\documentclass[fleqn,usenatbib,twocolumn]{mnras}

\usepackage[T1]{fontenc}

\DeclareRobustCommand{\VAN}[3]{#2}
\let\VANthebibliography\thebibliography
\def\thebibliography{\DeclareRobustCommand{\VAN}[3]{##3}\VANthebibliography}

\usepackage{graphicx}	
\usepackage{amsmath}	
\usepackage{amssymb}	
\usepackage{multirow}	

\newcommand{\tariq}[1]{}
\newcommand\HIbold{$\textbf{H}\scriptstyle\mathbf{I}$}
\newcommand{\w}{W_{50}}
\newcommand{\wobs}{W_{50,\text{obs}}}
\newcommand{\ewobs}{\delta\wobs}

\newcommand{\msun}{\,\textrm{M}_{\odot}}
\newcommand{\mbar}{M_{\text{bar}}}
\newcommand{\kms}{\, \text{km}\,\text{s}^{-1}}
\newcommand{\vmax}{V_{\text{max}}}
\newcommand{\vflat}{V_{\text{flat}}}
\newcommand{\overlap}{$\mathcal{O}$}
\newcommand{\improvement}{$\mathcal{I}$}

\newcommand{\mvir}{M_{\text{vir}}}
\newcommand{\rhi}{r_{\text{HI}}}

\newcommand{\HI}{\ensuremath{\mathrm{H}\scriptstyle\mathrm{I}}}

\newcommand{\lw}{$W_{50}$}

\setkeys{Gin}{keepaspectratio,width=\columnwidth}

\def\tightlist{} 
\usepackage[section]{placeins}
\makeatletter
\AtBeginDocument{%
  \expandafter\renewcommand\expandafter\subsection\expandafter{%
    \expandafter\@fb@secFB\subsection
  }%
}
\makeatother

\makeatletter
\AtBeginDocument{%
  \expandafter\renewcommand\expandafter\subsubsection\expandafter{%
    \expandafter\@fb@secFB\subsubsection
  }%
}
\makeatother

\usepackage{newtxtext,newtxmath}

\title[The haloes of HI-selected galaxies]{Inferring dark matter halo properties for \HIbold-selected galaxies}

\author[T. Yasin et al.]
{Tariq Yasin,$^{1}$\thanks{\href{mailto:tariq.yasin@physics.ox.ac.uk}{tariq.yasin@physics.ox.ac.uk}} 
Harry Desmond,$^{1,2,3}$
Julien Devriendt,$^{1}$
and Adrianne Slyz$^{1}$
\\ 
$^{1}$Astrophysics, University of Oxford, Denys Wilkinson Building, Keble Road, Oxford, OX1 3RH, UK\\
$^{2}$McWilliams Center for Cosmology, Department of Physics, Carnegie Mellon University, 5000 Forbes Ave, Pittsburgh, PA 15213, USA\\
$^3$Institute of Cosmology \& Gravitation, University of Portsmouth, Dennis Sciama Building, Portsmouth, PO1 3FX, UK\\
}

\date{Accepted XXX. Received YYY; in original form ZZZ}

\pubyear{2015}

\begin{document}
\label{firstpage}
\pagerange{\pageref{firstpage}--\pageref{lastpage}}
\maketitle

\begin{abstract}
We set constraints on the dark matter halo mass and concentration of \textasciitilde22,000 individual galaxies visible both in \HI{} (from the ALFALFA survey) and optical light (from the SDSS). This is achieved by combining two Bayesian models, one for the \HI{} line width as a function of the stellar and neutral hydrogen mass distributions in a galaxy using kinematic modelling, and the other for the galaxy's total baryonic mass using the technique of inverse subhalo abundance matching. We hence quantify the constraining power on halo properties of spectroscopic and photometric observations, and assess their consistency. We find good agreement between the two sets of posteriors, although there is a sizeable population of low-line width galaxies that favour significantly smaller dynamical masses than expected from abundance matching (especially for cuspy halo profiles). Abundance matching provides significantly more stringent bounds on halo properties than the \HI{} line width, even with a mass--concentration prior included, although combining the two provides a mean gain of 40\% for the sample when fitting an NFW profile. We also use our kinematic posteriors to construct a baryonic mass--halo mass relation, which we find to be near power-law, and with a somewhat shallower slope than expected from abundance matching. Our method demonstrates the potential of combining photometric and spectroscopic observations to precisely map out the dark matter distribution at the galaxy scale using upcoming \HI{} surveys such as the SKA.
\end{abstract}




\hypertarget{introduction}{%
\section{Introduction}\label{introduction}}

In the \(\Lambda\text{CDM}\) paradigm, galaxies form from the condensation of baryons in the centre of dark matter haloes, virialised overdensities seeded by primordial perturbations in the density field at the end of the inflationary epoch \citep{moGalaxyFormationEvolution2010}. Elucidating the ``galaxy--halo connection'', the set of correlations between the properties of galaxies and those of their host dark matter haloes, is a key endeavour in modern astrophysics, vital both to construct theoretical models of galaxy formation within \(\Lambda\text{CDM}\) and to enable tests of the model itself that rely on the
dark matter distribution on small or large scales.

The first evidence for dark matter (DM) came from the dynamics of the Coma cluster \citep{zwickyRotverschiebungExtragalaktischenNebeln1933}, and the kinematics of astrophysical objects remains one of our most powerful probes of its distribution. The most detailed modern studies of this type use radio interferometry to measure the rotation curves (RCs) of galaxies, the rotational velocity of the gas as a function of radius (e.g.~\citealt{walterTHINGSHINearby2008,lelliSPARCMassModels2016}). 
This has allowed detailed comparison between different halo models~\citep{katzTestingFeedbackmodifiedDark2017, liComprehensiveCatalogDark2020}, provided insights into the scaling relationships between galaxies (e.g.~\citealt{lelliBaryonicTullyFisherRelation2019, desmondUncorrelatedVelocitySize2019}) and afforded tests of modified gravity~ \citep{burrageRadialAccelerationRelation2017,naikConstraintsChameleonGravity2019,chaeTestingStrongEquivalence2021}. However, due to the observational expense, even the largest collations of RCs only contain hundreds of galaxies, with ongoing surveys such as MIGHTEE-HI \citep{maddoxMIGHTEEHIHIEmission2021} and APERTIF \citep{oosterlooLatestApertif2010a} aiming to extend this to thousands. 

In contrast, the width of the global \HI{} 21-cm emission line of a galaxy (henceforth \lw) is a readily observable kinematic tracer and has already been derived for tens of thousands of galaxies. The current state of the art is the ALFALFA survey \citep{haynesAreciboLegacyFast2018}, which has produced a catalogue of \lw{} for \textasciitilde30,000 galaxies out to $z\sim0.06$ using a blind drift scan with the ARECIBO telescope. Future single dish and interferometric surveys such as WALLABY (\(5 \times 10^5\) galaxies; \citealt{koribalskiWALLABYSKAPathfinder2020}) and the SKA ($\sim10^9$ galaxies out to $z\sim2$; \citealt{yahyaCosmologicalPerformanceSKA2015}) will increase the number of \lw{} observations by orders of magnitudes while extending to significantly higher redshift.

The drawback of using \lw{} rather than resolved RCs to localise dark matter is that it lacks spatial information within the galaxy, instead providing a weighted average across the RC. In particular, \lw{} is dependant on both the spatial distribution of \HI{} gas in the galaxy and the velocity profile that traces the total gravitational potential.
If the \HI{} gas disc extends into the flat part of the RC with speed $V_{\textrm{flat}}$, the \HI{} spectral profile takes the form of the classic symmetric two-horned shape, due to flux building up at the wavelengths equivalent to the Doppler shifting of the \HI{} line by \(\pm V_{\textrm{flat}}\). In this case the inclination-corrected line width $W_{50}/\sin{i}$ is approximately \(2V_{\textrm{flat}}\), which is typically
dominated by the gravity of the dark matter halo. However, depending on the \HI{} distribution, \lw{} can either be weighted towards the inner parts of the galaxy where the baryonic contribution is important, causing it to be closer to twice the peak velocity \(V_{\text{max}}\), or to lower values if the rotation curve is still rising or observed close to edge-on, causing a single peaked profile. \lw{} can also be affected by morphological asymmetries caused by environmental interactions or internal processes \citep{bokEnhancedProfileAsymmetries2019,reynoldsAsymmetriesLVHISVIVA2020,wattsGlobalAsymmetriesIllustrisTNG2020}. Despite these complexities \lw{} contains a large amount of potential information about the dynamical mass of a galaxy. \citet{lelliBaryonicTullyFisherRelation2019} find that for the SPARC sample \lw{} from archival single-dish data give a tighter Baryonic Tully--Fisher Relation than all velocity summary statistics (\(V_{\max}\), \(V_{2R_e}\),\(V_{2.2}\)) other than \(V_{\text{flat}}\).

In this study we extract constraints on halo properties for individual galaxies with \lw{} measurements. To do this we develop a Bayesian forward-modelling framework: we first construct models for the full \HI{} line profile of a galaxy for given baryonic and dark matter mass distribution and then compress this to the \lw{} statistic. Unlike models that simply equate \lw{} to some summary statistic of the rotation curve (typically $\vflat$ or $\vmax$), possibly with empirical corrections for the effects described above, our method naturally takes full account of the \HI{} distribution and shape of the rotation curve
based on the available observational data. We test the accuracy of our method, which uses survey data and empirical relationships to construct the distribution of baryons, using the resolved spectroscopy and photometry of the SPARC data set.

The properties of dark matter haloes can also be inferred from photometric data using empirical models that connect the properties of haloes in dark matter-only (DMO) cosmological simulations to large-scale galaxy surveys. These models assign galaxy properties to simulated haloes by making simple parameterised assumptions, which are then tested and the parameters constrained using observations such as galaxy clustering (see \citealt{wechslerConnectionGalaxiesTheir2018} for a review). This avoids invoking a full galaxy formation model as in hydrodynamical simulations or semi-analytical models, the complex baryonic physics of which are not yet well understood. We will use a particular empirical model called subhalo abundance matching (SHAM) to provide further constraints on the dark matter distribution of our observational sample, and to test for consistency between inferences made from photometric and spectroscopic data. 

SHAM assigns galaxy masses based on the virial mass or rotation velocity of their halo, with some scatter to allow for stochasticity in the galaxy--halo connection \citep{kravtsovDarkSideHalo2004,conroyModelingLuminosityDependentGalaxy2006,behrooziComprehensiveAnalysisUncertainties2010a,guoHowGalaxiesPopulate2010,mosterConstraintsRelationshipStellar2010}. The ALFALFA survey has revealed that \HI{}-selected samples are much more weakly clustered than optically-selected \citep{liClusteringGalaxiesFunction2012,martinCLUSTERINGCHARACTERISTICSISELECTED2012,papastergisClusteringALFALFAGalaxies2013}, and recent work has shown that reproducing their clustering signal requires performing SHAM on a subset of the simulated halo population selected on properties that have a similar clustering bias, such as formation time \citep{guoConstrainingHaloMass2017, stiskalekDependenceSubhaloAbundance2021}. 

SHAM, along with the similar empirical technique of Halo Occupation Distribution modelling, are commonly combined with the halo model or DMO simulations to construct mock catalogues for the massive volumes of the sky probed by cosmological surveys. They can also be applied in the inverse direction: using photometric observations of galaxies to map out the dark matter distribution \citep[e.g.][]{desmondReconstructingGravitationalField2018}. Therefore it is vital that these empirical methods are validated by independent tests, especially at low masses where clustering constraints become weak or unavailable.

In this work we use the SHAM prescription of  \citet{stiskalekDependenceSubhaloAbundance2021} (henceforth ST21), which is specifically tailored to the \HI{}-selected galaxies of the ALFALFA sample, to construct an inverse SHAM method which produces a Bayesian posterior on the halo properties of a galaxy given its baryonic mass as observational input. Comparing the independent constraints on halo properties from \HI{} kinematics and SHAM will allow us to answer the following questions:

\begin{itemize}
\tightlist
\item
  What is the DM content of \HI{}-selected galaxies and how does this correlate with galaxy variables at both the population and individual galaxy levels?
\item
  To what extent are the halo properties implied by kinematics and photometry
  consistent?
\item
  How much information on the DM distribution is contained in cheaper photometric observations of a galaxy, versus the more expensive \lw?
\item
  Does combining kinematic and abundance matching posteriors give us tighter constraints on halo properties than either one alone?
\end{itemize}

The paper is structured as follows. Section~\ref{data} describes the observed and simulated data we use. Section~\ref{methods} describes our Bayesian models and inference for \lw{} and SHAM, and the metrics we use to characterise the posteriors of individual galaxies. In Section~\ref{results-secresults} we present the constraints on halo properties obtained by the two models for our sample of \HI{}-selected galaxies, their summary metrics and their correlation with galaxy properties. We expand on the implications of our results, discuss potential systematic errors and compare to the literature in Section~\ref{discussion}, and conclude in Section~\ref{conclusions}. We assume \(H_0 = 70 \kms\), and all logarithms are base-10.

\hypertarget{data}{%
\section{Observed and simulated data}\label{data}}

\hypertarget{sec:obs_data}{%
\subsection{ALFALFA}\label{sec:obs_data}}

We take
\HI{} line widths, \HI{} masses and distances from the ALFALFA\footnote{\url{http://egg.astro.cornell.edu/alfalfa/data/index.php}}  \citep{haynesAreciboLegacyFast2018} data set. This contains \HI{} line sources in 7000 \(\text{deg}^2\) of high Galactic latitude out to a redshift of 0.06 generated from a blind survey using the Arecibo telescope.
The complete \(\alpha.100\) catalogue contains \textasciitilde31500 extragalactic sources with \HI{} masses ranging from \(10^6\) to \(10^{11}\) \(\textrm{M}_{\odot}\). We use the \(W_{50}\) line width measurement, which is the velocity width measured at 50\% of the peak flux. This is calculated
by identifying the two peaks of the classic twin horned \HI{} line profile, then fitting polynomials to the data between the peak and zero flux on each side. \(W_{50}\) is corrected for instrumental broadening, as described in \citet{springobDigitalArchiveHI2005}, but not for turbulent motion, disk inclination or cosmological effects. The catalogue also contains \(W_{20}\), the velocity width measured at 20\% of the peak flux. Theoretically this is expected to be a better tracer of the dark matter-dominated outskirts of the galaxy. However \citet{haynesAreciboLegacyFast2018} find the measurement less robust, and its errors harder to quantify, and so only an error on \(W_{50}\) is provided.

The catalogue also contains distances. For objects with \(cz > 6000\) km/s this is \(cz\), where \(z\) is the redshift in the CMB frame measured from the centroid of the \HI{} line. For objects closer than this a distance is calculated using the local peculiar velocity model of \citet{mastersCosmologyVeryLocal2005}. Primary distances are used if they are available in the literature. The \HI{} mass \(M_{\text{HI}}\) is provided, calculated from the total integrated \HI{} flux and the assumed distance. A code indicates the reliability of each detection based on the signal-to-noise ratio (SNR) and other observational criteria. Code 1 sources (of which there are 25,434) are the highest-quality observations,
whilst code 2 sources may have lower SNR, but have been successfully crossmatched with known optical counterparts at the same redshift as the \HI{} line. We use both in our fiducial analysis.

\citet{yuStatisticalAnalysisProfile2022} perform a reanalysis of the ALFALFA raw spectra, including calculating a different, integrated definition of the line width using the "curve of growth" method of \citet{yuDeterminationRotationVelocity2020}. Their measurement (henceforth $W_{\text{Yu85}}$) is defined as the velocity width enclosing 85\% of the total flux. We use $W_{\text{Yu85}}$ in a non-fiducial model to test for possible systematics due to the ALFALFA data reduction.

\hypertarget{nasa-sloan-atlas}{%
\subsection{NASA-Sloan Atlas}\label{nasa-sloan-atlas}}

We use the NASA-Sloan Atlas (NSA) to obtain stellar masses and optical axis ratios (used to derive inclinations) for the ALFALFA galaxies. The NSA consists of images and parameters of local galaxies derived from Sloan Digital Sky Survey (SDSS) imaging data, with the addition of \emph{Galaxy Evolution Explorer} (GALEX) \citep{martinGalaxyEvolutionExplorer2005} data. We use NSA v1\_0\_1\footnote{\url{https://www.sdss.org/dr13/manga/manga-target-selection/nsa/}}  which is based on SDSS DR13 \citep{albareti13thDataRelease2017a} and contains \textasciitilde640,000 galaxies out to redshift \(z=0.15\). The image analysis pipeline utilises enhanced object detection, deblending and other improvements over standard SDSS processing, which improves its performance for larger and brighter galaxies \citep{blantonImprovedBackgroundSubtraction2011b}. The catalogue contains both elliptical Petrosian and S\'{e}rsic aperture photometry fits, with the former considered more reliable.
The fits are K-corrected to \(z=0.0\) using the \texttt{kcorrect code v4\_2} \citep{blantonKCorrectionsFilterTransformations2007}, which also estimates the stellar mass. A Chabrier initial mass function (IMF) is assumed and spectral energy distributions (SEDs) are fitted to the broadband optical SDSS fluxes as well as ultraviolet fluxes from \emph{GALEX} when available.

To crossmatch the NSA catalogue with ALFALFA we follow ST21 by requiring a 5 arcsecond skymatch between NSA galaxies and the ALFALFA$\times$SDSS optical counterparts of \citet{durbalaALFALFASDSSGalaxyCatalog2020}, and a maximum line-of-sight distance of 10 Mpc. These criteria should yield a low number of mismatches. There are 21,776 galaxies in our final ALFALFA$\times$NSA catalogue. We note that the $\text{ALFALFA}\times\text{SDSS}$ catalogue of \citet{durbalaALFALFASDSSGalaxyCatalog2020} contain estimates of stellar mass for \textasciitilde30,000 galaxies, but we prefer to use NSA's improved image analysis pipeline, which also improves consistency with ST21. 

\hypertarget{sparc}{%
\subsection{SPARC}\label{sparc_data}}

SPARC\footnote{\url{http://astroweb.cwru.edu/SPARC/}} \citep{lelliSPARCMassModels2016} is a database of 175 late-type galaxies for which both high quality \HI{} rotation curves and near-infrared Spitzer photometry are available. We use SPARC's publicly available resolved rotation curves, detailed mass models and \lw{} measurements, as well as the \HI{} surface density as a function of radius (F. Lelli, private communication) in our analysis. The high resolution SPARC data is used as a truth against which we test the approximations that allow us to model the much larger ALFALFA data set. 

The sample spans a large range in luminosity (\(10^7\) to \(10^{12} L_{\odot}\)), surface brightness (\textasciitilde5 to \textasciitilde{}\(5000 \: L_{\odot} \text{pc}^{-2}\)), gas mass (\textasciitilde{}\(10^7\) to \(~10^{10.6} M_{\odot}\)) and morphology (S0 to Im/BCD). The advantage of Spitzer photometry is that at $3.6 \mu\text{m}$ the mass-to-light ratio of galaxies is relatively constant, which helps to break the degeneracy between the velocity contribution from the stars and the dark matter. We perform a crossmatch between ALFALFA and SPARC using the same procedure as for the NSA, which yields 45 matches.

\citet{lelliBaryonicTullyFisherRelation2019} compile \HI{} line widths from various sources in the literature and for various definitions for the SPARC galaxies. The closest line width definition they include to ALFALFA's \lw{} is \(W_{\textrm{50Mc}}\): the width at 50\% flux of the mean flux, where the mean flux is taken across the whole line width, with corrections for instrumental resolution and relativistic broadening. As we find the two measures are very similar for galaxies that have both,
we test our \lw{} model using the 125 galaxies in SPARC for which there is either an ALFALFA \(W_{50}\) or a SPARC \(W_{\textrm{50Mc}}\). 

\hypertarget{simulation-data}{%
\subsection{Simulation data}\label{simulation-data}}

As input to our SHAM model we use the 140 Mpc/\(h\) "Shin-Uchuu" box of the Uchuu suite of cosmological N-body simulations\footnote{\url{http://skiesanduniverses.org/Simulations/Uchuu/}} \citep{ishiyamaUchuuSimulationsData2021}. Shin-Uchuu contains \(6400^3\) particles, with a particle mass of \(8.97 \times 10^5 M_{\odot}/h\) and force softening length of \(0.4  h^{-1} \text{kpc}\). The simulation uses the GreeM N-body code \citep{ishiyamaGreeMMassivelyParallel2009,ishiyama2015} and the 2018 Planck flat \(\Lambda \text{CDM}\) cosmology \citep{planckcollaborationPlanck2018Results2020}: \(H_0=67.74 \kms \, \text{Mpc}^{-1}\); \(\Omega_{\text{0}}=0.3089\); $\lambda_0=0.6911$, scalar spectral index \(n_{s} = 0.9667\); root-mean-square matter fluctuation on 8 Mpc/\(h\) scales \(\sigma_8 = 0.8159\). Haloes and subhaloes were identified using the Rockstar halo finder \citep{behrooziRockstarPhaseSpaceTemporal2013} and the Consistent Trees Merger Tree Code \citep{behrooziGRAVITATIONALLYCONSISTENTHALO2013}. The halo finder calculates the halo mass in a profile-independent manner by simply summing the mass of the particles belonging to each halo using the overdensity condition \(\Delta_{\text{vir}}=178\) of  \citet{bryanStatisticalPropertiesXray1998}. We use the trimmed Shin-Uchuu catalogue of haloes with $\mvir>10^9\msun$, so each halo has a minimum of 1100 particles and is therefore well-resolved.
We use the same overdensity condition to define virial quantities for the haloes in our kinematic model.

\hypertarget{methods}{%
\section{Methodology}\label{methods}}

\subsection{Overview and verification with SPARC}\label{kin_overview}

For each galaxy in our sample we have two separate Bayesian models: the kinematic model, for which the observable is the ALFALFA \(W_{50}\), and the SHAM model, for which the observable is the galaxy's baryonic mass \(\mbar\). Our kinematic model also contains nuisance parameters describing the galaxy, the priors of which are informed by the NSA and ALFALFA data. The free parameters and their priors for each model are listed in Table \ref{tab:parameters}. 

\begin{table*}
\caption{\label{tab:parameters}The free parameters in our abundance matching and kinematic models, their physical definitions and their Bayesian priors. All parameters are sampled in logarithmic space except inclination and distance.}
\begin{tabular}{ |c|c|c|c|c| }
  \hline
   & \textbf{Parameter} & \textbf{Units} & \textbf{Definition} & \textbf{Prior} \\
  \hline

  \raisebox{-.5\normalbaselineskip}[0pt][0pt]{\rotatebox[origin=c]{90}{\textbf{AM}}} & $M_\text{vir}$ & $M_{\odot}$ &  Virial mass & \multirow{2}{*}{ 2D halo probability density from Uchuu simulation} \\
   & $c_{0.1}$ &  & Custom halo concentration, defined in Eq.~(\ref{eq:conc}) & \\

  \hline

  \hline
   & $M_{\text{vir}}$ & $M_{\odot}$ & Virial mass $M_{\text{vir}} = M_{\text{halo}} + M_{\text{bar}}$ (see Section~\ref{kin_overview}) & Flat in range $\log(\mbar/\msun) < \log(M_{\text{vir}} / \msun) < 15.5$ \\
   & $c_{0.1}$ &   & Custom halo concentration, defined in Eq.~(\ref{eq:conc}) & Flat in range $0.5 < \log c_{0.1} < 2$, or halo mass--concentration relationship\\
   & $M_{\text{HI}}$ & $M_{\odot}$ & \HI{} mass & Gaussian prior from ALFALFA observed value and its uncertainty \\
  \raisebox{-.5\normalbaselineskip}[0pt][0pt]{\rotatebox[origin=c]{90}{\textbf{Kinematics}}}& $M_{*}$ & $M_{\odot}$ & Stellar mass & Gaussian prior from NSA observed value with adopted 0.2 dex uncertainty \\
   & $r_{\text{HI}}$ & kpc  & Scale length of exponential gas disc & Gaussian prior from \cite{wangNewLessonsHI2016} mass--size relationship\\
   & $r_{\text{disk}}$ & kpc & Scale length of exponential stellar disc & Gaussian prior from empirical relationship  based on $M_*$ \citep{duttonDarkHaloResponse2011c}\\
   & $r_{\text{bulge}}$  & kpc & Half light radius of stellar bulge & Gaussian prior from empirical relationship based on $M_*$  \citep{duttonDarkHaloResponse2011c} \\
   & $D$ & Mpc & Physical distance to galaxy & Gaussian prior from ALFALFA value and its uncertainty\\
   & $i$ & deg & Inclination ($0^\circ $ face on; $90^\circ$ edge on)  & Equation (\ref{eq:inclin}) with NSA $b/a \pm 10\% $, and a flat prior on $q$ in the range 0.15 to $b/a$.  \\
  \hline

\end{tabular}
\end{table*}

DMO simulations (on which our SHAM model is based) implicitly assume that
baryons and dark matter behave identically and hence are perfectly mixed. The virial halo masses listed in their halo catalogues (which we label $\mvir$) are therefore different to the masses inferred from fitting parameterised halo profiles to kinematic data (which we label $M_{\text{halo}}$) according to $\mvir = M_{\text{halo}} + \mbar$ where $\mbar$ is the mass of \textit{observed} stars and cold gas. $M_{\text{halo}}$ therefore includes DM as well as any unobserved baryons, which are inferred indiscriminately in the kinematic analysis. We assume that the sum of these components follows a standard density profile (Section~\ref{dark-matter}), e.g. because they have the same distribution.
In reality, the distribution of hot gas around galaxies and its association with the dark matter is poorly known. \cite{hafenOriginsCircumgalacticMedium2019} found that $f_{\text{b}}$ can vary substantially between galaxies, as part of the circumgalactic medium can be accreted from the intergalactic medium and other galaxies. However, this potential systematic uncertainty in $\mvir$ of up to $\sim15\%$ (for a halo completely stripped of baryons) is still small compared to the typical uncertainties on our derived halo masses. We do not modify halo concentrations to account for the fact that a fraction of $M_{\text{halo}}$ is in baryons (see also Sec.~\ref{halo-mismatch-concentration}).

In our kinematic model we sample \(\log{M_{\text{vir}}} = \log(M_{\text{halo}} + \mbar)\) rather than $M_\text{halo}$ itself, setting
the lower bound on its flat prior to be $\log\mbar$. We find some galaxies to have non-zero posterior probability at \(M_{\text{halo}}=0\), so this parameterisation aids sampling by raising the lower limit of the posterior to a finite value. We could perform the SHAM inference with the same parameterisation as the kinematics, but it is unnecessary. Our SHAM posterior does not approach \(M_{\textrm{halo}}=0\), so we can simply convert to or from $M_{\text{vir}}$ in post-processing. 

\begin{figure*}
\hypertarget{fig:schema}{%
\centering
\includegraphics[width=\textwidth]{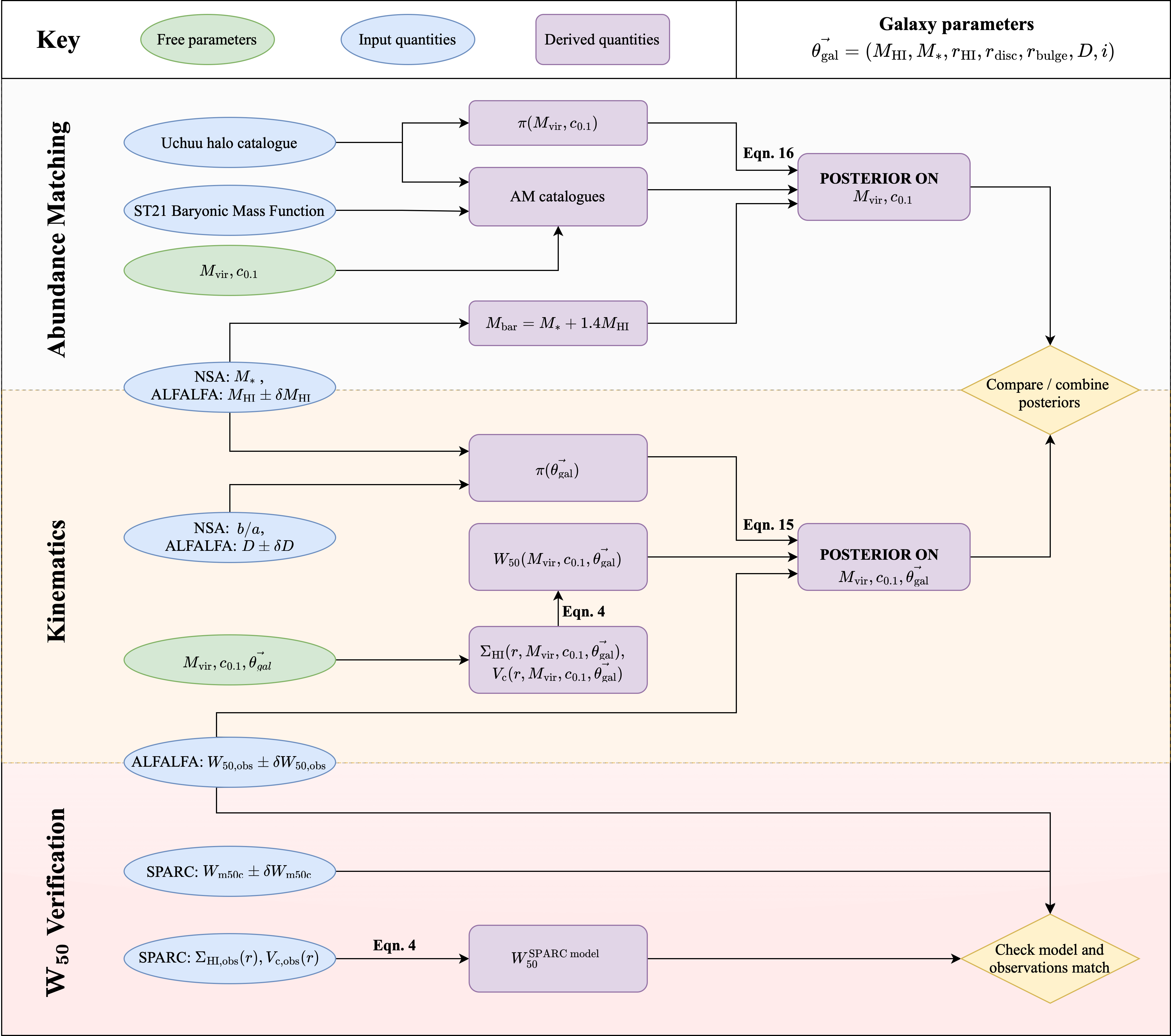}
\caption{Schematic of our workflow for constraining halo mass and concentration from abundance matching and kinematics.
The kinematics observable is the ALFALFA \lw. Observations of galaxy parameters from the NSA and ALFALFA inform the priors on the free parameters in the kinematic inference. The observed \(M_{\textrm{HI}}\) and \(M_*\) are also used to calculate the galaxy's baryonic mass \(M_\text{bar}\), which is the observable in the subhalo abundance matching model. The resolved rotation curves ($V_\text{c,obs}$) and \HI{} surface density observations ($\Sigma_{\text{HI,obs}}$) from the SPARC data set are used to verify that the model for calculating \lw{} (equation~\ref{eq:convolution}) can match observations.}\label{fig:schema}
}
\end{figure*}

In Fig.~\ref{fig:schema} we show relationship between our observed and simulated input quantities and our Bayesian models, as well as the verification of the \lw{} model using the SPARC data.
In Section~\ref{sec:kinematic_section} we describe our kinematic model and its verification using SPARC,
and in Section~\ref{sec:abundance_matching} our SHAM model.

\hypertarget{sec:kinematic_section}{%
\subsection{Kinematic model}\label{sec:kinematic_section}}

The kinematic model puts constraints on halo parameters by combining a parameterised form for the halo density profile with a baryonic mass distribution to forward-model \lw{} for comparison with the ALFALFA data. The forward-model consists of three steps:

\begin{enumerate}
\def\labelenumi{\arabic{enumi}.}
\tightlist
\item
  Calculate the model rotation curve $V_{\text{c}}(r)$  and \HI{} surface density $\Sigma_{\text{HI}}(r)$ for the galaxy for a given set of model free parameters.
\item
  Construct the model \HI{} line profile from $\Sigma_{\text{HI}}(r)$  and $V_{\text{c}}(r)$.
\item
  Calculate \(W_{50}\) from the line profile and compare to the observed value.
\end{enumerate}

As SPARC has mass models for the \HI{} and stars of each galaxy, as well as the observed RC, we can use it to carry out checks on our model:

\begin{enumerate}
\item
  Are the observed $W_{50}$ of real galaxies well modelled as a product of their azimuthally averaged RC and \HI{} surface density?
\item
  How much do our simple mass models based on empirical relations and non-resolved observations bias line widths compared to the detailed mass models in SPARC?
\item
  How accurate are optical inclinations from SDSS photometry compared to those derived from tilted-ring fits to spatially resolved velocity fields?
\end{enumerate}

The \HI{} spectrum is a spatially integrated quantity, so to calculate \lw{} we must first model the full RC. We assume the rotational speed is equal to the circular speed (we discuss the validity of this in Sec.~{\ref{rotation-curve-not-reflected-in-line-width}}). The total circular speed, \(V_\text{c}(r)\), is the sum in quadrature of the circular speed due to each of the galaxy's components (dark matter, stellar bulge, stellar disc, gas disc), 


\begin{equation}\protect\hypertarget{eq:ph}{}{
V_{\mathrm{c}}^{2}(r)=V_{\mathrm{DM}}|V_{\mathrm{DM}}|+ V_{\mathrm{bulge}}|V_{\mathrm{bulge}}|  +  V_{\mathrm{disc}}|V_{\mathrm{disc}}|
 + V_{\mathrm{gas}}|V_{\mathrm{gas}}|,
}
\end{equation}
where $V_x$ is also a function of $r$. To construct the \HI{} spectrum from $V_{\text{c}}(r)$ and $\Sigma_{\text{HI}}(r)$ we follow the method of \citet{obreschkowSIMULATIONCOSMICEVOLUTION2009}, by considering the gas disc as a series of flat rings each with a constant circular speed $V_\text{c}$.
The flux of the \HI{} spectrum at a given wavelength $\lambda$ corresponds to \HI{} gas with radial velocity $V_{\lambda}$, which we define relative to the galaxy's systemic velocity so the midpoint of the spectrum is at  $V_{\lambda}=0$. The flux at wavelength $\lambda$ due to a single, infinitely thin ring of gas 
is given by
\begin{equation}
\tilde{\psi}\left(V_{\lambda}, V_{\mathrm{c}}\right)=\left\{\begin{array}{ll}
\frac{1}{\pi \sqrt{V_{\mathrm{c}}^{2}-V_{\mathrm{\lambda}}^{2}}} & \text { if }\left|V_{\mathrm{\lambda}}\right|<V_{\mathrm{c}} \\
0, & \text { otherwise }
\end{array}\right.
\end{equation}

The singularity is smoothed by introducing a velocity dispersion \(\sigma_{\HI{}}\) for the gas, which models its random motion. We fix the dispersion to be 10 km/s, based on observations of nearby galaxies \citep{leroyStarFormationEfficiency2008a,mogotsiHICOVelocity2016}.
This broadens the flux from the ring:
\begin{equation}
\psi\left(V_{\mathrm{\lambda}}, V_{\mathrm{c}}\right)=\frac{\sigma_{\HI{}}^{-1}}{\sqrt{2 \pi}} \int^{+\infty}_{-\infty} d V \exp \left[\frac{\left(V_{\mathrm{\lambda}}-V\right)^{2}}{-2 \sigma_{\HI{}}^{2}}\right] \tilde{\psi}\left(V, V_{\mathrm{c}}\right).
\end{equation}
The total flux observed at $\lambda$ is then obtained by integrating over the galaxy:
\begin{equation}\protect\hypertarget{eq:convolution}{}{
\Psi_{\mathrm{HI}}\left(V_{\lambda}\right)=\frac{2 \pi}{M_{\mathrm{HI}}} \int_{0}^{\infty} dr \: r \Sigma_{\mathrm{HI}}(r) \psi\left(V_{\lambda}, V_{\mathrm{c}}(r)\right).
}\label{eq:convolution}\end{equation}
As our model profiles are symmetrical, $W_{50}$ is calculated by identifying the peak flux and then finding the outermost point where the flux is 50\% of the maximum.

\begin{figure*}
\hypertarget{fig:spectrum}{%
\centering
\includegraphics[width=\textwidth]{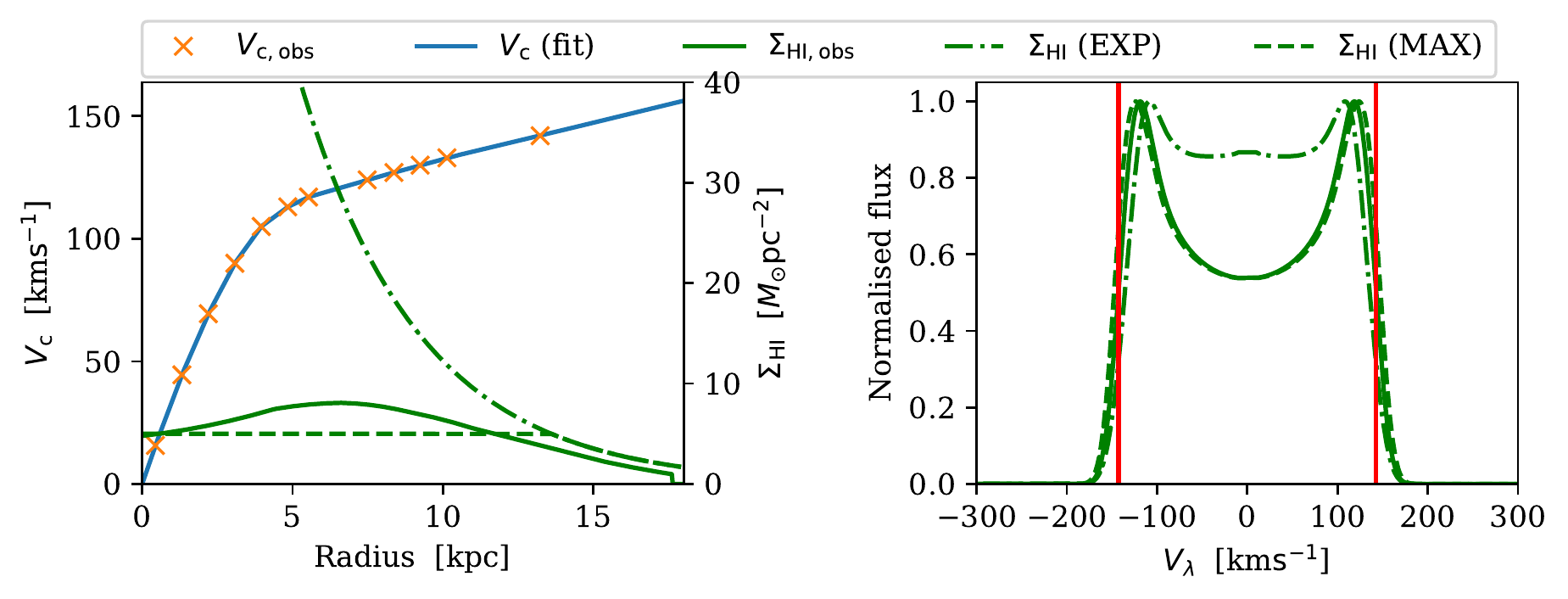}
\caption{\emph{Left panel}: For the low surface brightness galaxy F568-1 we plot $V_\text{c,obs}$ from SPARC (orange) and its fit (blue) with linear extrapolation. The observed $\Sigma_{\text{HI,obs}}$ from SPARC is shown, as well as the untruncated exponential model (EXP) and the truncated model (MAX) with $\Sigma_{\text{MAX}}=5 M_{\odot}\text{pc}^{-2}$ (see Section~{\ref{baryons}}). The model $\Sigma_{\text{HI}}$ are based on the mass-size relationship of \citet{wangNewLessonsHI2016}.
\emph{Right panel}: The \HI{} spectra produced by applying equation (\ref{eq:convolution}) to $V_{c}$ (fit) and the three $\Sigma_{\HI{}}$. $V_{\lambda}$ is the observed radial velocity relative to the galaxy's systemic velocity. The resulting \lw{} are: $285.4\kms$ for $\Sigma_{\text{HI,obs}}$ (shown in red); $271.2\kms$ for EXP; $287.3\kms$ for MAX. The untruncated model produces lower values of \lw{} generally. This galaxy is an example where even though the MAX model does not provide a perfect fit to $\Sigma_{\HI{}}$, it is still capable of reproducing \lw.}}
\label{fig:spectrum}
\end{figure*}

To check the accuracy of modelling \lw{} in this way, we calculate a ``SPARC model $W_{50}$'' by applying equation (\ref{eq:convolution}) to $V_\textrm{c,obs}$, the observed inclination-corrected RC, and $\Sigma_{\textrm{HI,obs}}$, the observed \HI{} surface density profile, of the SPARC galaxies. For some SPARC galaxies the \HI{} surface density observations extend beyond the final data point of the RC. We extrapolate the RC as $V_{\text{flat}}$ if it is defined for the galaxy \citep{lelliSMALLSCATTERBARYONIC2015}, or linearly extrapolate the RC if it is not. We show this procedure for a single galaxy in Fig.~\ref{fig:spectrum}. To approximate uncertainties on the SPARC model $W_{50}$ we use the maximum and minimum \lw{} generated by combinations of the observational errors on the RC and $V_{\text{flat}}$, and extrapolating linearly or with $V_{\text{flat}}$.

\begin{figure}
\hypertarget{fig:model_residuals}{%
\centering
\includegraphics{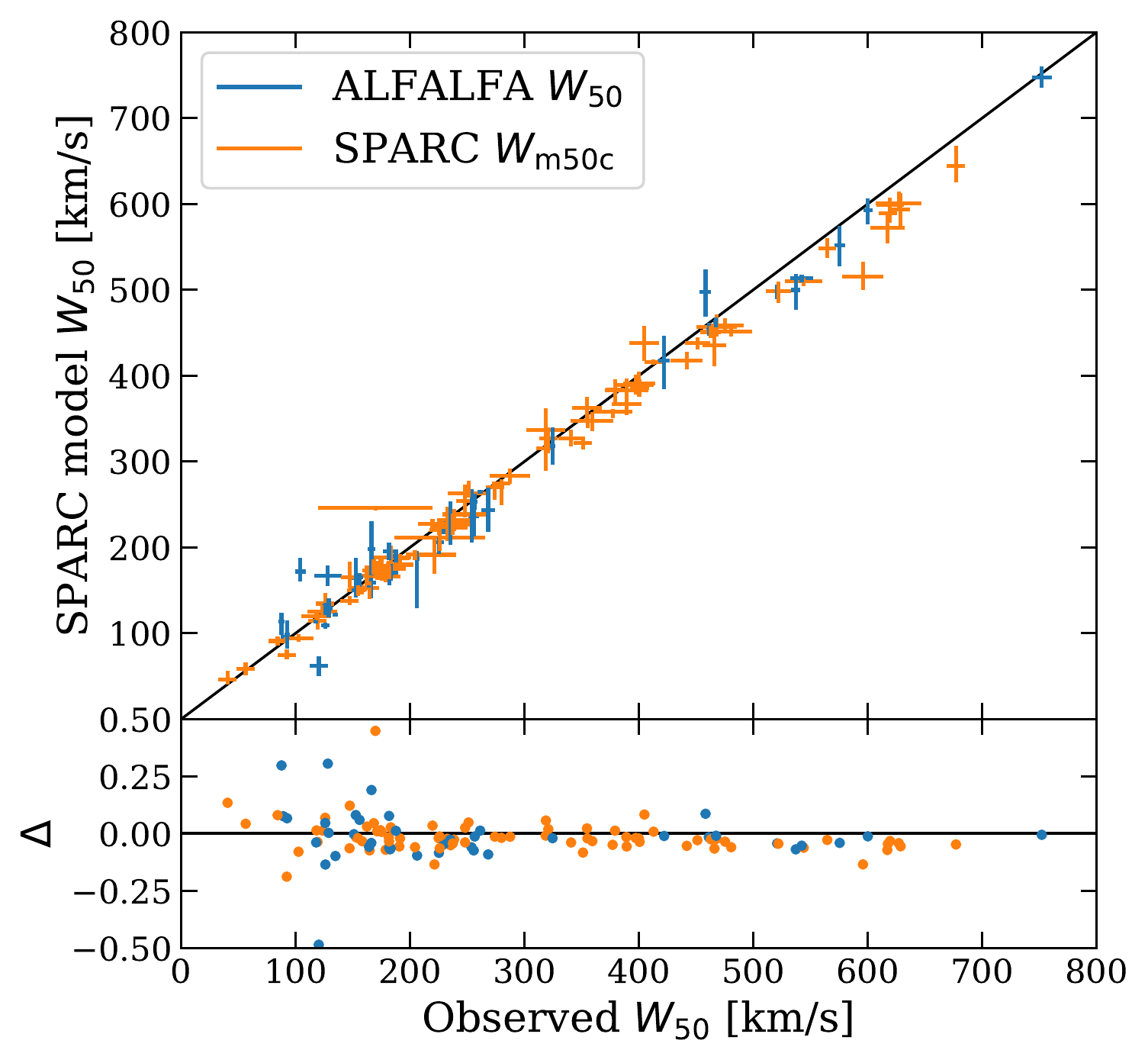}
\caption{The SPARC model \lw{} (calculated from the observed SPARC \HI{} surface density and RC using equation \ref{eq:convolution}) plotted against the observed \lw, for the 125 galaxies in SPARC with either an ALFALFA \lw{} or SPARC $W_{50\text{Mc}}$ (see Section~\ref{sparc_data}). The subplot shows the relative differences $\Delta=\frac{\text{Model}-\text{Observed}}{\text{Observed}}$. The model and observations are in good agreement, although there is increasing scatter in the residuals at lower values. Both quantities are calculated/corrected using the inclinations from the SPARC kinematic fitting.}\label{fig:model_residuals}
}
\end{figure}

In Fig.~\ref{fig:model_residuals} we compare the SPARC model \lw{} to the observed \lw{} for the 125 galaxies in the SPARC data set with either an ALFALFA \lw{} or the very similar SPARC \(W_{\textrm{50Mc}}\) (described in Section~\ref{sparc_data}). Although the uncertainties on the SPARC model \lw{} are crude, there are only 3 galaxies for which there is \(>3\sigma\) tension between the model and observed value. It is interesting that the model \lw{} on average slightly underpredicts the observed value across a large range, suggesting it cannot be due to a constant effect such as instrumental broadening. A possible cause is non-circular motions such as outflows.
We conclude that the model works well for \(W_{50} \gtrsim 200 \kms\). The scatter between model and observations increases at lower \(W_{50}\), but this can be explained by the increased uncertainty in the extrapolation of the RC. 

To apply the \lw{} model to the ALFALFA galaxies, we now need to construct model $\Sigma_{\HI}(r)$ and $V_{\text{c}}(r)$ for the ALFALFA sample.

\hypertarget{dark-matter}{%
\subsubsection{Dark matter distribution}\label{dark-matter}}

The circular speed contribution from the DM is calculated by assuming a halo profile, with the halo mass \(M_{\text{halo}}\) and its concentration as free parameters. There are a large number of different halo profiles in literature (see e.g. \citealt{liComprehensiveCatalogDark2020}).
We perform our analysis for the cuspy Navarro-Frenk-White (NFW) \citep{navarroUniversalDensityProfile1997} and the cored Burkert \citep{burkertStructureDarkMatter1995} profiles as representatives of cusped and cored profiles respectively. We caution that neither of these is likely to be entirely accurate after accounting for baryonic feedback, as we discuss further in Section~\ref{sec:kinematic_caveats}.

\textbf{NFW:} NFW found that the haloes in cosmological DMO simulations are well fit by a universal density profile

\begin{equation}
\rho_{\mathrm{NFW}}(r)=\frac{\rho_{s}}{\left(\frac{r}{r_{s}}\right)\left[1+\left(\frac{r}{r_{s}}\right)\right]^{2}},
\label{eq:nfw}\end{equation}
where $r_s$ is a scale radius and $\rho_{s}$ a characteristic density. The enclosed mass at radius \(r\) is
\begin{equation}{
M_{\mathrm{NFW}}(r) =4 \pi \rho_s r_{s}^{3}\left[\ln (1+x)-\frac{x}{1+x}\right]
,}\end{equation}
where \(x\equiv r/r_s\).

\textbf{Burkert:} The Burkert profile was proposed as a modification to the PISO profile in order to improve the fit to observations of dwarf spheroids at larger radii. 
The density profile is
\begin{equation}\protect\hypertarget{eq:burkert}{}{
\rho_{\text {Burkert }}(r)=\frac{\rho_s}{\left(1+\frac{r}{r_{s}}\right)\left[1+\left(\frac{r}{r_{s}}\right)^{2}\right]},
}\label{eq:burkert}\end{equation}
and the enclosed mass is given by
\begin{equation}\protect\hypertarget{eq:ph}{}{
M_{\text {Burkert }}(r)=2 \pi \rho_s r_{c}^{3}\left[\frac{1}{2} \ln \left(1+x^{2}\right)+\ln (1+x)-\arctan (x)\right].
}\end{equation}

It should be noted that whilst differing at small radii, at large radii both NFW and Burkert profiles have the same slope \(\rho \propto 1/r^3\). Therefore the large extrapolation necessary to calculate the halo's total mass (the virial radius lies far outside the radius probed by \lw) is similar for both. The contribution of the halo to the circular speed at radius \(r\) is \(V_{\text{DM}} = \sqrt{GM_{\text{DM}}(r)/r}\).
It is most convenient to calculate relative to the virial quantities

\begin{equation}\protect\hypertarget{eq:ph}{}{
\frac{V_{\text{DM}}(r)}{V_{\text{halo}}} = \sqrt{\frac{M_{\text{DM}}(r)}{M_{\text{halo}}} \frac{R_{\text{halo}}}{r}}.
}\end{equation}
where \(V_{\text{halo}} \equiv \sqrt{G M_{\mathrm{halo}}/R_{\mathrm{halo}}}\) is the circular speed at the virial radius $R_{\text{halo}}$ (which is inferred from $M_{\text{halo}}$).

Traditionally, concentration is defined as
\begin{equation}\protect\hypertarget{eq:ph}{}{
c_{\text{halo}} = \frac{R_{\text{halo}}}{r_{-2}},
}\end{equation}
where \(r_{-2}\) is the radius at which the slope of the logarithmic density profile is equal to -2 (for NFW \(r_s=r_{-2}\)). We instead use a new concentration definition

\begin{figure}
\hypertarget{fig:conc}{%
\centering
\includegraphics{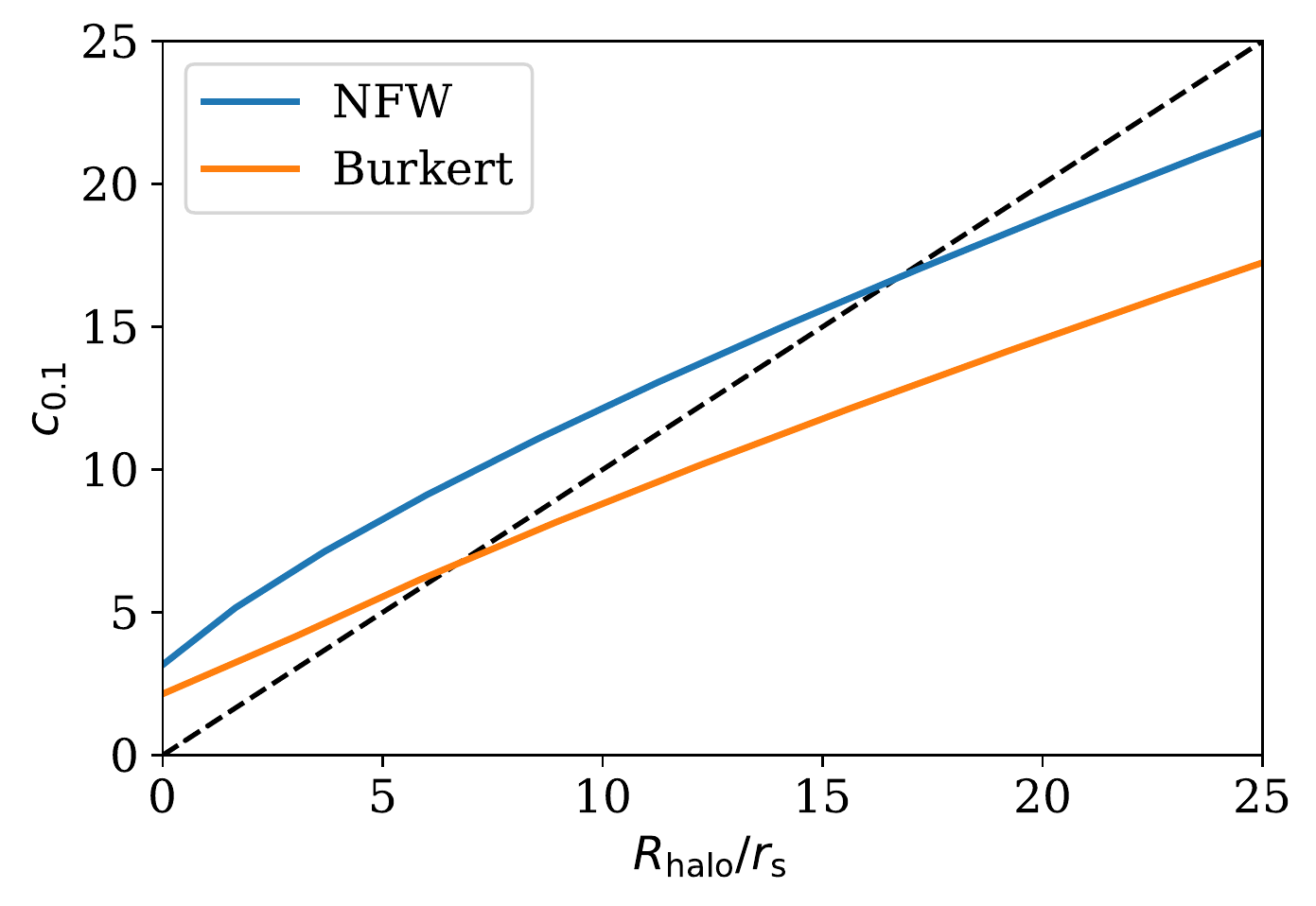}
\caption{The mapping between $R_{\text{halo}}/r_{s}$ (where $R_{\text{halo}}$ is the virial radius) and our new concentration definition $c_{0.1}$ (equation~\ref{eq:conc}). The scale length $r_s$ is defined separately for the NFW (equation~\ref{eq:nfw}) and Burkert (equation~\ref{eq:burkert}) profiles. For the NFW profile $r_{s}=r_{-2}$.}\label{fig:conc}
}
\end{figure}

\begin{equation}\protect\hypertarget{eq:conc}{}{
c_{0.1} = \frac{R_{\text{halo}}}{r_{0.1}},
}\label{eq:conc}\end{equation}
where \(r_{0.1}\) is the radius enclosing 10\% of the halo mass. This new definition has three advantages: 1) unlike \(r_{-2}\), \(r_{0.1}\) is defined for all halo profiles; 2) \(r_{0.1}\) can be calculated from haloes in simulations without assuming a profile but just counting particles in spheres grown from the centre; 3) it is easier to interpret, as it does not depend on the profile shape. In Fig.~\ref{fig:conc} we show the mapping between \(c_{0.1}\) and $R_{\text{halo}}/r_s$, which we calculate numerically.

\hypertarget{baryons}{%
\subsubsection{Baryon distribution}\label{baryons}}

We set the prior on $M_*$ to be a Gaussian with mean equal to the NSA $M_*$ and a scatter of 0.2 dex for all galaxies, as a representative value of uncertainty on the mass-to-light ratio.
We compare the NSA $M_*$ to the GALEX-SDSS-WISE Legacy Catalog \citep{salimDustAttenuationCurves2018} values; this is a smaller sample than NSA, but all galaxies have SDSS spectra and WISE photometry which are used in their alternative SED fitting pipeline. For the \textasciitilde11,000 galaxies in both catalogues and ALFALFA, the mean difference in $M_*$ is 0.04 dex, with a standard deviation of 0.19 dex. With 0.2 dex uncertainty, the NSA $M_*$ are also consistent with those from SPARC (adopting $\Upsilon_*=0.5 \msun/L_{\odot}$ as per \citealt{lelliSPARCMassModels2016}). We scale the NSA $M_*$ according to the distances listed in ALFALFA.

We use $M_{\textrm{HI}}$ and its uncertainty from ALFALFA. \citet{mcgaughScalingRelationsMolecular2020} find that the molecular gas mass for late-type galaxies is around 7\% of the stellar mass.  
This correction is minor for most of our sample. Therefore to be consistent with ST21 we set the total baryonic mass to be simply \(\mbar = M_* + 1.4M_{\text{HI}}\), where the factor of 1.4 accounts for cosmological helium and metals. 

To estimate the spatial distribution of the stars we use the empirical relationships of \citet{duttonDarkHaloResponse2011c}, who use the GIM2D software \citep{simardDEEPGrothStrip2002} to perform two-component bulge and disc fits to $r$-band and $g$-band images and hence derive structural properties of \textasciitilde200,000 late and early type galaxies from SDSS DR7. They find empirical relationships for $M_*-r_{\text{disc}}$ and $M_*-r_{\text{bulge}}$, and the bulge fraction. We use their relationships for late-type galaxies, as these comprise the bulk of our sample. We fix the bulge fraction to the mean relationship and set the Bayesian priors on $r_{\text{disc}}$ and $r_{\text{bulge}}$ to be a Gaussian with mean given by \citet{duttonDarkHaloResponse2011c} and a scatter of 0.5 dex (estimated from their fig. 4).

For the model \HI{} surface density $\Sigma_\text{HI}(r)$ we use the empirical relationship of \citet[henceforth W16]{wangNewLessonsHI2016}, based on a sample of 500 mainly late-type galaxies with spatially resolved \HI{} observations. They find 

\begin{equation}\protect\hypertarget{eq:size_mass}{}{
\log \left(D_{\HI} / \mathrm{kpc}\right)=0.51 \log \left(M_{\mathrm{\HI}} / M_{\odot}\right)-3.29,
}\label{eq:size_mass}\end{equation}
with 0.06 dex of scatter, where \(D_{\text{HI}}\) is the diameter of the \(1 M_{\odot} / \text{pc}^2\) isophote of $\Sigma_{\HI{}}$. 
W16 also find that for late-type galaxies, there is a homogeneous exponential profile in the outer parts with scale length $r_{\HI{}} =0.1 D_{\text{HI}}$. Adopting this and equation (\ref{eq:size_mass}) specifies the full $\Sigma_{\text{HI}}(r)$. We also fit exponential profiles to the outer radii of the SPARC galaxies, and find a similar relationship. 

Observed \HI{} discs are not exponential all the way to the centre \citep{leroyStarFormationEfficiency2008a,wangObservationalTheoreticalView2014,wangNewLessonsHI2016}.
Many galaxies have $\HI{}$ cores or holes as the high density \HI{} converts to $\text{H}_2$. The conversion is also dependant on metallicity and temperature, so varies between galaxies \citep{bigielUniversalNeutralGas2012}. Therefore we truncate $\Sigma_{\text{HI}}(r)$ to have a maximum value \(\Sigma_{\textrm{max}}\), which is set by the requirement to reproduce the observed $M_{\text{HI}}$. For the W16 model this gives \(\Sigma_{\textrm{max}} \approx 5 \msun\), which is consistent with observations. This truncated model is better at handling slowly rising RCs, where an unrealistic central peak in \HI{} could overweight the velocities at low radius and thus cause \lw{} to be underestimated. However it is not very sensitive to the precise value of \(\Sigma_{\textrm{max}}\). Fig.~\ref{fig:spectrum} compares the model $\Sigma_{\HI}(r)$ to SPARC observations for a particular galaxy. Although the model is not a precise match, we see that this does not cause a large shift in \lw{} for a galaxy with a slowly rising RC. This is because the \HI{} density towards the centre changes the flux and shape, but does not shift the position of \(W_{50}\), which is mainly set by the outer, flatter part of the RC that forms the horns.

\begin{figure}
\hypertarget{fig:inc_residuals}{%
\centering
\includegraphics{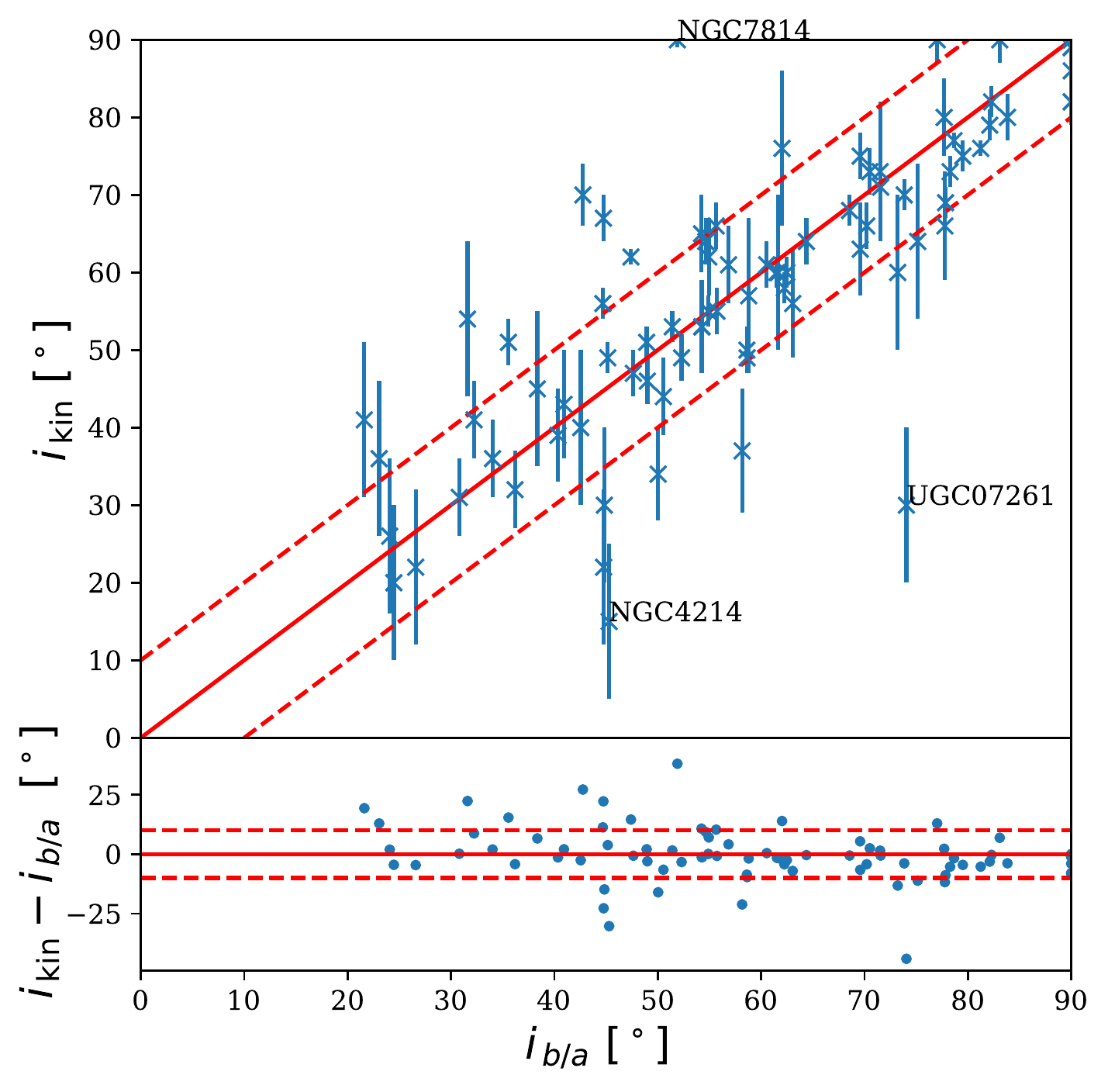}
\caption{The inclinations from NSA $r$-band optical axis ratios assuming $q=0.2$ ($i_{\, b/a}$) plotted against inclinations obtained from tilted ring fits to the resolved \HI{} velocity field ($i_{\text{\,kin}}$) for the SPARC sample. The solid red line shows equality, and the dashed red lines \(\pm 10^\circ\) disagreement. The subplot shows their residuals. Our fiducial model has a flat prior on $q$, so galaxies above the solid red line are accounted for in the scatter of the posterior probability on halo properties. We account for galaxies below the red line by adopting a $10\%$ uncertainty on $b/a$.
}\label{fig:inc_residuals}
}
\end{figure}

\hypertarget{inclination-secinclination}{%
\subsubsection{\texorpdfstring{Inclination }{Inclination \{Section~\}}}\label{sec:inclination}}

The model circular speed $V_{\text{c}}(r)$ must be corrected for inclination (\(i=0 ^\circ\) face on); \(i=90 ^\circ\) edge on) by a factor of $\sin(i)$. As ALFALFA does not resolve the \HI{} disc, inclination is derived from optical photometry. Assuming axisymmetry, the relationship between intrinsic relative thickness $q$, observed axis ratio \(b/a\) and inclination is
\begin{equation}\protect\hypertarget{eq:inclin}{}{
\cos ^{2} i=\frac{(b / a)^{2}-q^{2}}{1-q^{2}}.
}\label{eq:inclin}\end{equation}
Typically \(b/a\) is obtained from infrared photometry, due to the lower extinction. However this light predominantly comes from older stars that do not reside in a thin disc. There is much discussion on $q$ in the literature, and its variation with galaxy type. It is common to simply assume \(q=0.2\) for line width surveys \citep{zwaanVelocityFunctionGasrich2010}, based on population studies of late-type galaxies (e.g. \citealt{unterbornInclinationDependentExtinction2008}). 
Based on colour, bulge fractions and S\'{e}rsic index the majority of the ALFALFA sample are late-type. However as baryonic mass increases (above $\sim 10^{10} \msun$), there is an increasing population of galaxies in ALFALFA with properties more consistent with early-type galaxies. Dwarf galaxies are also observed to have thicker discs \citep{mendez-abreuWhichGalaxiesHost2010}.

In view of the importance of, and uncertainty in, inclinations we consider three different models for it, ordered from least to most conservative:

\begin{enumerate}
\def\labelenumi{\arabic{enumi}.}
\item
  Assume q=0.2 and derive $i$ from equation (\ref{eq:inclin}) using the observed $b/a$ and its uncertainty $\sigma_{b/a}$.
\item
  Give $q$ a flat prior between 0.15 (the lowest $b/a$ in the NSA) and the observed $b/a$ of the galaxy. Then  the prior on inclination is $\pi(i | b/a \pm \sigma_{b/a} )$, as calculated from equation (\ref{eq:inclin}). We use this prior in our fiducial analysis as it is more conservative. However, assuming the true distribution of $q$ for our sample is peaked at lower values, it will bias our results towards higher $i$ and hence lower dynamical mass. This is because $b/a \le 1$, so changing $q$ has a bigger effect on the numerator of Eq.~\ref{eq:inclin} than the denominator.
\item
  Simply assume the galaxies are randomly oriented. The angle between two random vectors in 3D is distributed as \(\sin(i)\), so the prior on inclination is then \(\pi(i) = \frac{1}{2}\sin(i)\).
\end{enumerate}

For \(b/a\) we use the SERSIC\_BA value from the NSA, which is calculated from single-component two-dimensional S\'{e}rsic fits to $r$-band photometry. In Fig.~\ref{fig:inc_residuals} we plot the SPARC kinematic inclinations against the NSA optical inclinations (with $q=0.2$), for the 84 galaxies in both SPARC and the NSA. The kinematic inclinations are obtained from tilted-ring fits to the velocity fields, and so are expected to be more accurate.

There are a couple of extreme outliers. The optical inclination of UGC\,07261 severely overpredicts the kinematic inclination. On inspection, its SDSS image shows it to be heavily barred, causing it to have a far lower $b/a$ that does not reflect the stellar disc itself, which appears to be nearly face on. This could be mitigated by calculating $b/a$ using outer isophotes only.
However we find all other measures of $b/a$ in the NSA, such as those based on Petrosian fits at radii containing 50\% and 90\% of the total light, produce less plausible inclinations across the whole sample.

The optical inclination of NGC\,7814 is a severe underestimate of the kinematic value. The SDSS image reveals it to be an edge-on thin disk with an extremely prominent bulge, resulting in a high $b/a$. The flat prior on $q$ in our fiducial model accounts for this, as $q=1$ is appropriate for a bulge. In fact our fiducial model is flexible enough that it can account for any case where the optical inclination is an underprediction of the true value.

NGC\,4214 is the most severe example of a galaxy where the inclination is overpredicted by the optical $b/a$ due to the irregularity of the light distribution in its SDSS image. Irregularity has a tendency to make circular objects appear to have a lower $b/a$. To account for these galaxies, we adopt an uncertainty of $10\%$ on $b/a$, which makes the SPARC kinematic and optical inclinations consistent within $2\sigma$ for all galaxies where the optical $b/a$ overpredicts inclination, except for UGC\,07261.

\hypertarget{likelihood}{%
\subsubsection{Likelihood}\label{likelihood}}

We compare our model line width to the observed line width by using Bayes' theorem to calculate the probability for the parameters \(\theta\) given model \(\mathcal{M}\) and data $D$,

\begin{equation}
\mathcal{P}(\theta |D,\mathcal{M})=\frac{\mathcal{P}(D|\theta,\mathcal{M}) \mathcal{P}(\theta|\mathcal{M})}{\mathcal{P}(D|\mathcal{M})}.
\end{equation}
We choose the likelihood to be


\begin{equation}\protect\hypertarget{eq:ph}{}{
\mathcal{P}(\wobs|\theta,\mathcal{M}) = \frac{\exp\{ 
-(\wobs - \w)^2 / (2\ewobs^2) \}}
{\sqrt{2\pi}\ewobs} }.
\end{equation}
We also repeat the same inference, but imposing the mass--concentration relation (converted to $c_{0.1}$) derived from the Uchuu simulations \citep[][equation 2]{ishiyamaUchuuSimulationsData2021} as a prior, with 0.11 dex log-normal scatter. This breaks the degeneracy between mass and concentration.

\begin{figure}
\centering
\includegraphics{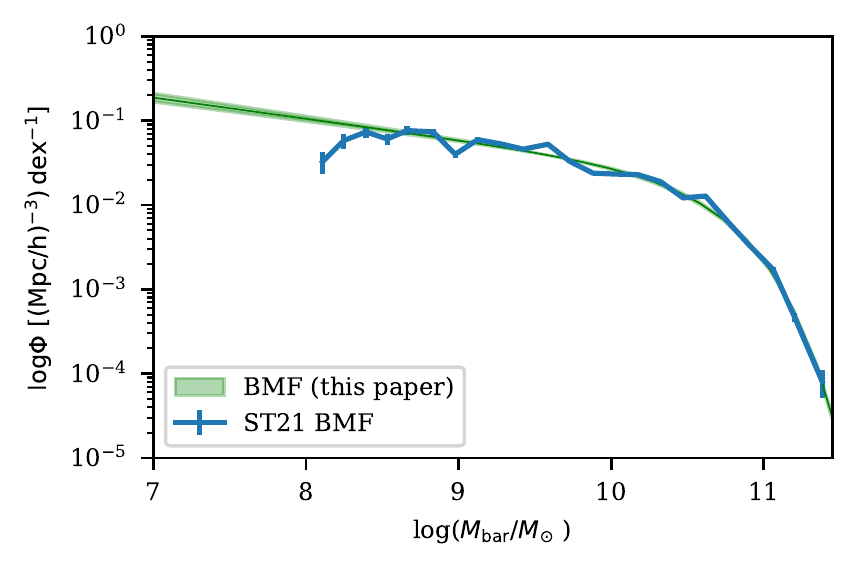}
\caption{The baryonic mass function (BMF) used in the abundance matching analysis (green line), derived by fitting a Schechter function to the ST21 BMF (blue). The four lowest mass data points are not included in the fit, as the faint end is potentially biased by incomplete treatment of selection effects (ST21). The shaded band shows the $1\sigma$ uncertainty.}\label{fig:bmf}
\end{figure}

\hypertarget{sec:abundance_matching}{%
\subsection{Abundance matching}\label{sec:abundance_matching}}

In its simplest form, SHAM posits that the most massive (or brightest) galaxy forms in the most massive halo or subhalo, the second most massive galaxy in the second most massive halo and so on \citep{kravtsovDarkSideHalo2004}. When applied to a galaxy survey and an equivalently sized simulation box this yields a monotonic relationship between halo mass and galaxy mass. \citet{conroyModelingLuminosityDependentGalaxy2006} showed that this simple non-parametric model could produce an excellent fit to galaxy clustering from the present day up to $z=5$. The model has been extended to allow stochasticity through the SHAM scatter parameter $\sigma_{AM}$, which models both intrinsic scatter in the galaxy--halo connection and scatter from observational uncertainties in the galaxy mass or luminosity \citep[][henceforth BCW10]{behrooziComprehensiveAnalysisUncertainties2010a}. Halo assembly bias can be included in the SHAM framework by allowing secondary halo parameters such as concentration to affect the order in which galaxies are assigned to haloes \citep{reddickConnectionGalaxiesDark2013, lehmannConcentrationDependenceGalaxyHalo2016, chaves-monteroSubhaloAbundanceMatching2016,stiskalekDependenceSubhaloAbundance2021}.

In SHAM the property used to rank galaxies is traditionally luminosity or stellar mass, as these quantities are readily available for the large samples required to calculate accurate abundance and correlation functions. In this work we use a SHAM model from ST21 (section 4.2) that is based on a sample of HI-selected galaxies, and instead ranks galaxies using their baryonic mass. This is possible as the galaxy sample is a crossmatch of NSA and ALFALFA, and hence contains both a stellar mass and \HI{} mass for each galaxy. Baryonic mass is expected to be more fundamentally related to halo mass than stellar mass as star formation has only an indirect effect on galaxies' baryonic mass fractions.
This is implied empirically by the Tully--Fisher relation and radial acceleration relation, both of which are tighter and more regular when plotted in terms of total cold baryonic mass rather than stellar mass.

\begin{figure}
\centering
\includegraphics{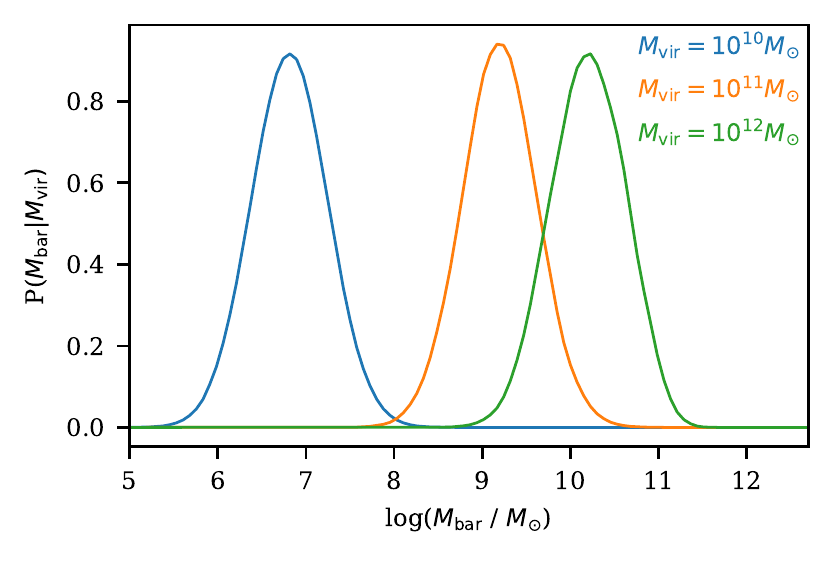}
\caption{Probability distributions $P(\mbar|\mvir)$ for different values of $\mvir$, calculated from an ensemble of abundance matching catalogues. These form the likelihood in our Bayesian inverse abundance matching method (Section \ref{sec:abundance_matching}).}\label{fig:example_pdfs}
\end{figure}

\begin{table*}
\caption{\label{tab:statistics}The statistics we use to characterise and compare the 2D posteriors in halo mass and concentration from kinematics and AM.} \begin{tabular}{ |c|p{8cm}|p{8cm}| }
  \hline
  \textbf{Metric} & \textbf{Definition} & \textbf{Interpretation} \\
  \hline
   $\mathcal{F}$ & 
   The fraction of the marginalised 1D $\log(M_{\textrm{vir}})$ posterior probability from kinematics that lies in the region $\log(M_{\textrm{vir}}) < \log(M_{\textrm{bar}}) + 0.2$. &
    The extent to which a galaxy is compatible with having $M_{\textrm{vir}} = M_{\text{bar}}$ (i.e. $M_{\text{halo}}=0$). Galaxies with large observational uncertainties will have higher $\mathcal{F}$. We interpret $\mathcal{F} > 0.01$ as $M_{\text{halo}}=0$ not being excluded by \lw.\\
   \hline
   $\mathcal{O}$ &
   The fraction of the marginalised 2D $\{ M_{\textrm{vir}},c_{0.1} \}$ posterior probability from kinematics that lies inside the $2\sigma$ contour of the SHAM posterior.  &
   The level of agreement between SHAM and kinematics. We interpret \overlap{} $< 0.01$ as the two models being in tension. \\
   \hline
   $\mathcal{I}$ &
   $\left[ \cfrac{\sigma_{\text{AM}}}{\sigma_{\text{AM+KIN}} }- 1\right]$, where $\sigma_{\text{AM}}$ is the size of the $1\sigma$ contour of the marginalised 1D $M_{\text{vir}}$ posterior for AM, and $\sigma_{\text{AM+KIN}}$ is the same quantity for the combined posterior of SHAM and kinematics.  &
   The improvement in the constraint on $M_{\textrm{vir}}$ obtained when combining kinematics with SHAM, compared to SHAM alone. $\mathcal{I}=0$ corresponds to no improvement and $\mathcal{I}=1$ to a constraint that is twice as tight.\\

  \hline

\end{tabular}
\end{table*}

To construct the galaxy baryonic mass function (BMF) we fit a Schechter function \citep{schechterAnalyticExpressionLuminosity1976}
\begin{equation}
\phi\left(M_{\mathrm{bar}}\right)=\ln (10) \phi_*\left(\frac{M_{\mathrm{bar}}}{M_*}\right)^{\alpha+1} e^{-\left(\frac{M_{\mathrm{bar}}}{M_*}\right)},
\end{equation}
where the fit parameters are $M_*,\phi_*$ and $\alpha$, to the ALFALFA$\times$NSA BMF of ST21 (plotted in their fig.~2). We remove the four lowest mass data points, as ST21 suggest that the faint end is biased by incomplete treatment of selections effects. The derived BMF, which we plot in Fig.~\ref{fig:bmf}, has parameters $\alpha=-1.24 \pm 0.02$, $\log(M_* h^2 /M_{\odot})=10.20 \pm 0.02$ and $\phi^*=3.3 \pm 0.2 \times 10^{-3} h^3 \text{Mpc}^{-3} \mathrm{dex}^{-1}$ (with $\alpha$ and $\phi_*$ anti-correlated with $M_*$). The uncertainty on the BMF (which is dominated by $\alpha$) does not significantly impact the assignment of galaxies to haloes in the mass range probed by ALFALFA, and so is not propagated through our analysis. We find the fitted value of $\alpha$ is in good agreement with that of the ALFALFA \HI{}MF derived by \citet{jonesALFALFAMassFunction2018}, which is unhampered by selection effects induced by the ALFALFA$\times$NSA crossmatch.

For the SHAM proxy, haloes are first selected that have a peak-mass redshift below a certain value \(z_{\text{cut}}\), and then ranked by their present-day $\mvir$. The  \(z_{\text{cut}}\) parameter allows the model to reproduce the weaker clustering of \HI{}-selected samples with lower \(\sigma_{\mathrm{AM}}\). ST21 found \(z_{\text{cut }}=0.22_{-0.2}^{+0.4}\) and \(\sigma_{\mathrm{AM}}=0.42_{-0.2}^{+0.8}\) dex from clustering constraints. We use these maximum likelihood values in our fiducial model. These uncertainties are also not propagated, as discussed further in Section~\ref{sec:am_caveats}.

The basic abundance matching procedure links each halo in the catalogue to a galaxy baryonic mass as follows: ~

\begin{enumerate}
\def\labelenumi{\arabic{enumi}.}
\tightlist
\item
  Deconvolve the galaxy mass function with the chosen SHAM scatter (BCW10).
\item
  Remove haloes with a peak formation time before the redshift \(z_{\text{cut}}\).
\item
  Rank haloes by the proxy (in our case \(\mvir\)).
\item
  Assign a baryonic mass to each halo by matching abundances as described above.
\item
  Add the SHAM scatter according to the prescription of BCW10.
\end{enumerate}
We repeat this process 500 times, thereby assigning 500 $\mbar$ to each halo in the catalogue. We bin the haloes onto a grid of $\mvir$ and $c_{0.1}$, and use kernel density estimation to calculate the 1D probability distribution \(P(\mbar|\mvir,c_{0.1})\) for each bin. The probability distributions are then interpolated over the entire space of \(\{\mvir,c\}\). As our proxy is simply $\mvir$ after $z_\text{cut}$ preselection, the probability is actually independent of $c_{0.1}$ here. We show examples in Fig.~\ref{fig:example_pdfs}. The posterior probability that the halo of a galaxy has parameters \(\{\mvir,c_{0.1}\}\), given that it is observed to have baryonic mass \(M_{\text{bar,obs}} \pm \delta M_{\text{bar,obs}}\), is

\begin{equation}\protect\hypertarget{eq:am_eq}{}{
P(\mvir,c_{0.1}|M_{\text{bar,obs}})= \frac{P(M_{\text{bar,obs}}|\mvir,c_{0.1}) P(M_{\text{vir}},c_{0.1})}{P(M_{\text{bar,obs}})}.
}\label{eq:am_eq}\end{equation}
The likelihood for the observed baryonic mass is connected to the likelihood for the ``true'' $\mbar$ from abundance matching by

\small \begin{equation}\protect\hypertarget{eq:ph}{}{
P(M_{\text{bar,obs}} |\mvir,c_{0.1}) = \int^{\infty}_{0} P(M_{\text{bar,obs}}|M_{\text{bar}}) P(M_{\text{bar}}|\mvir,c_{0.1}) \, dM_{\text{bar}},
}\end{equation}
\normalsize
where
\begin{equation}\protect\hypertarget{eq:ph}{}{
P(M_{\text{bar,obs}} |M_{\text{bar}}) =  \frac{1}{\sqrt{2 \pi} (\delta M_{\text{bar,obs}})} e^{-\frac{(M_{\text{bar,obs}} - M_{\text{bar}})^2}{2 (\delta M_{\text{bar,obs}})^2}}.
}\end{equation}
The prior \(P(M_{\text{vir}},c_{0.1})\) is the 2D probability density function in \{$\mvir,c_{0.1}$\} for all haloes in the simulation.

\hypertarget{sec:analysis}{%
\subsection{Inference methods and analysis of posteriors}\label{sec:analysis}}

\begin{figure*}
\hypertarget{fig:new_theory}{%
\centering
\includegraphics[width=13cm,height=15cm,keepaspectratio]{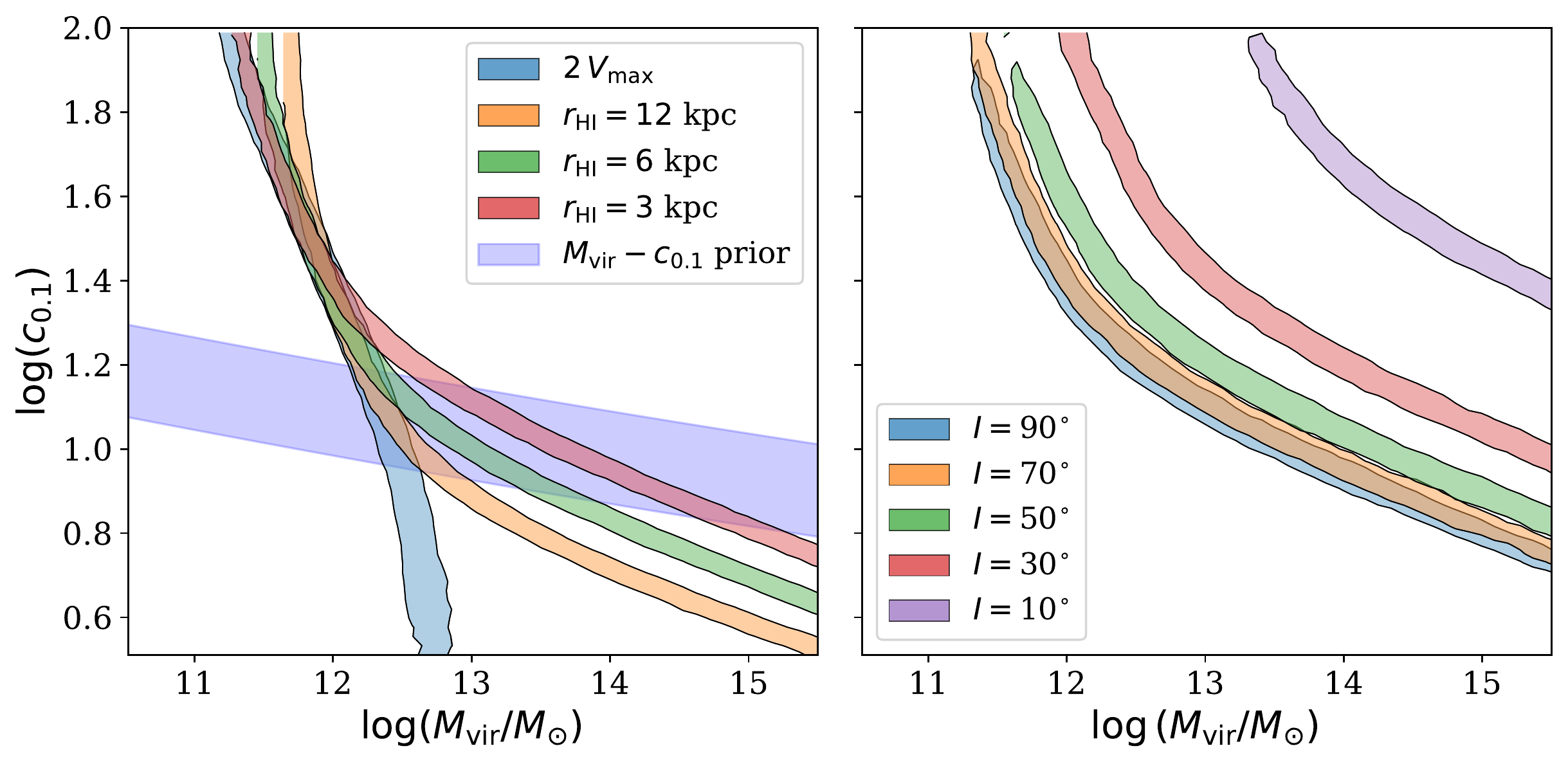}
\caption{The posterior for AGC 742791 (\(W_{50}=428 \pm 25 \, \kms\)), obtained by assuming the line width is only sourced by the DM, and with no uncertainty on inclination or the gas disc size \(r_{\text{HI}}\). The contours show the 1-sigma region, which has width only due to the uncertainty on the line width. In the left panel \(r_{\text{HI}}\) is varied. We also show the simple model where \(W_{50}=2\,V_{\text{max}}\), where $V_{\text{max}}$ is the maximum circular velocity of the dark matter halo, and the mass--concentration prior. As the size of the gas disc decreases and probes a smaller radius, the posterior deviates further from the \(2\,V_{\text{max}}\) model. In the right panel \(r_{\text{HI}}\) is fixed and the inclination \(i\) is varied.}\label{fig:new_theory}
}
\end{figure*}

We generate the posterior probability distributions using the \texttt{emcee} ensemble sampler \citep{foreman-mackeyEmceeMCMCHammer2013}. We initiate the sampler with 200 walkers and use the stretch move $a=2$. We run a small sample of galaxies until the strict \(\tau > 50\) autocorrelation time convergence condition is reached. For the whole sample we use only 10,000 steps, which for most galaxies does not reach $\tau = 50$, but we check our posteriors are unaffected using the galaxies with fully converged chains. We remove the first 5000 steps as burn-in. The acceptance fraction is \textasciitilde0.15. We rerun a subsample of galaxies with \texttt{Multinest} to check our results are not dependant on the choice of sampler. We place a flat prior on \(\log(M_{\text{vir}}/\msun)\) between \(\log(\mbar)\) and \(15.5\), corresponding to an extremely massive cluster. We sample in \(\log(c_{0.1})\), with a flat prior between a lower bound corresponding to the smallest value \(c_{0.1}\) can take for an NFW halo, and an upper bound of 2, as there are extremely few haloes in the Uchuu simulation with higher concentration. 

In Table~\ref{tab:statistics} we define a number of metrics to summarise and compare the posteriors produced by SHAM and kinematics. We quantify the extent to which the posterior is compatible with $M_{\text{halo}}=0$ using the $\mathcal{F}$ metric. It is defined as the fraction of the posterior for which $\log(M_{\text{vir}}) < \log(\mbar)+0.2$, with 0.2 chosen because it is the uncertainty on $M_*$, which is typically much larger than the uncertainties on $M_{\text{HI}}$ and so is roughly the maximum uncertainty on $\mbar$. 

The tension between the SHAM and kinematics posteriors is quantified by the the overlap metric \overlap{}, the fraction of the kinematics posterior that lies inside the $2\sigma$ contour of the SHAM posterior. We interpret two posteriors as being in tension if \overlap{}<0.01, roughly corresponding to a \(2\sigma\) disagreement. We choose 0.01 (rather than 0.05) so small posteriors that overlap with much larger posteriors are not considered to be in tension. This statistic is easy to calculate from the posteriors and can easily handle a variety of shapes of the kinematics posterior, at the price of a nontrivial interpretation. Like other assessments of tension such as the Bayesian evidence, this method is sensitive to the choice of prior. Two posteriors that overlap perfectly will also be considered more in tension if one of them is very wide. In this sense \overlap{} measures the similarity of the two posteriors along the lines of the Bhattacharyya distance \citep{bhattacharyyaMeasureDivergenceTwo1946}. We avoid using Bayes factors to evaluate tension, because commonly used samplers may not evaluate the evidence correctly for posteriors with plateaus and shallow likelihoods, which are common for our kinematic model (\citealt{schittenhelmPreprint2021,fowlieNestedSamplingCrosschecks2020}; although see \citealt{fowlieNestedSamplingPlateaus2021} for a solution). To ensure \overlap{} is calculated accurately we check that the $2\sigma$ contours have converged with respect to chain length, and we use a large number of samples (1,000,000) so the tails of the distributions are well resolved.

The metric \improvement quantifies the tightening of the constraint on $\mvir$ from combining kinematics with SHAM. It is defined relative to the SHAM constraint alone because this tends to be significantly tighter than the kinematic constraint, and because it is more common to have photometric than spectroscopic measurements.
We compare in $\mvir$ rather than $\mvir$--$c_{0.1}$, because $\mvir$ and $c_{0.1}$ are highly degenerate in the kinematic inference. We check that using a different contour, e.g.~\(2 \sigma\), or using the Kullback--Leibler divergence as a measure of information content, leads to the same conclusions. We redefine the SHAM likelihood to include the prior, so it can have the same flat prior on $\mvir$ and $c_{0.1}$ as the kinematic model. Then the combined posterior can be obtained by direct multiplication of the two separate posteriors, due to the assumption of independence.

\hypertarget{results-secresults}{%
\section{Results}\label{results-secresults}}

\hypertarget{individual-posteriors}{%
\subsection{Individual posteriors}\label{individual-posteriors}}

\subsubsection{Idealised dark matter-only inference}

 To understand the posterior shapes for the kinematic model, it is instructive to consider first the case of a single galaxy where we assume the potential is sourced purely by the dark matter, and the galaxy parameters ($i$, $M_*$, $r_{\text{HI}}$ etc.) are perfectly known. This leaves \(\mvir\) and \(c_{0.1}\) as the only free parameters and \(\delta W_{50}\) the only source of uncertainty. We apply this to AGC 742791 and show the posterior in Fig.~\ref{fig:new_theory} for different assumed values of \(\rhi\) and \(i\). The elongated posterior is due to the perfect degeneracy between mass and concentration; the same line width can be sourced by a high mass halo with a low concentration or vice versa. What matters to first order is the dynamical mass contained within the gas disc that contributes to the rotational velocity. As there are no additional parameters to marginalise over, the posteriors are the locus in the mass--concentration plane required to produce the observed \(W_{50}\), broadened by the observational error \(\delta W_{50}\). 

The naive model \(W_{50} = 2 \vmax\), where \(\vmax\) is the maximum circular speed of the halo, is only equal to our full model for values of \(c_{0.1}\) where \(\vmax\) occurs at a similar radius to \(\rhi\). When \(\vmax\) occurs at a larger radius than \(\rhi\) more halo mass is required to produce the same \(W_{50}\), so our model posterior curves to the right of the \(2\vmax\) posterior. This effect is more pronounced at low values of \(\rhi\) and \(c_{0.1}\), where \(\vmax\) occurs at a larger radius. Conversely, when \(\vmax\) occurs at a much smaller radius than \(\rhi\) there is a large contribution to the \HI{} spectral line from the gradually decreasing part of the rotation curve beyond \(V_{\textrm{max}}\), which biases \(W_{50}\) to lower values. More mass is required at a given concentration to produce the observed \(W_{50}\), so once again our posterior curves to the right of the \(2V_{\textrm{max}}\) case, with the effect more pronounced for higher values of \(\rhi\) and \(c_{0.1}\) (the effect is only significant at extreme $c_{0.1}$). Adding baryons to the idealised model would shift the posteriors to the left, as less DM is required to generate the observed \lw. 

The right hand panel of Fig.~\ref{fig:new_theory} shows the posterior for our idealised model at different assumed inclinations. At high inclination the \(\sin i\) correction to \(W_{50}\) is slowly changing, so as inclination decreases the posterior slowly sweeps to higher mass/concentration. As \(i\) decreases the posterior sweeps to higher mass and concentration at an increasing rate. Therefore our `randomly distributed' inclination model (3) would have a posterior with have a sharp peak at the \(i=90 ^{\circ}\) region, with a long tail all the way to the high \(\mvir\) and \(c_{0.1}\) prior boundary at low values of \(i\). For our fiducial inclination model (2), with a flat prior on intrinsic relative thickness \(q\), the minimum inclination occurs at $q=0.15$, where the inclination probability also peaks. Hence there is a competing effect between the peak in inclination probability at the lowest value of inclination, and the peak in the 2D posterior corresponding to \(i=90 ^\circ\) described above. For the simple $q=0.2$ model (1), the uncertainty on inclination is only due to the uncertainty on $b/a$, so the posterior is a band.

\subsubsection{The full model}

The output of our model for a single galaxy is the posterior probability distribution on the two halo parameters \(\mvir\) and \(c_{0.1}\) for both the SHAM and kinematic model, as well as the six galaxy parameters for kinematics (listed in Table~\ref{tab:parameters}). We also calculate the joint posterior of both models. Our kinematic model puts Gaussian priors on all galaxy parameters (except inclination), with the width set by the observational or empirical model uncertainty. The line width does not further constrain these galaxy parameters beyond their priors in most cases, so only the 2D and 1D posteriors on \(\mvir\) and \(c_{0.1}\) are of primary interest. Inclination $i$ is sometimes prevented from taking on very low values (face-on) if this would cause too high an intrinsic line width, requiring a dark matter halo with \(\mvir\) or \(c_{0.1}\) beyond the prior boundaries. It can also be prevented from taking on high values (edge-on), as for some galaxies this would make the intrinsic line width so low that the circular velocity from the baryons alone exceeds it. However the galaxy parameters are still important, as the uncertainty on them broadens the posteriors on halo properties, often dramatically. For example the high-end tail of the prior on $M_*$ can cause the $\mvir$ posterior to have a significant tail to low values.

\begin{figure*}
\hypertarget{fig:grid2d}{%
\centering
\includegraphics[width=\textwidth]{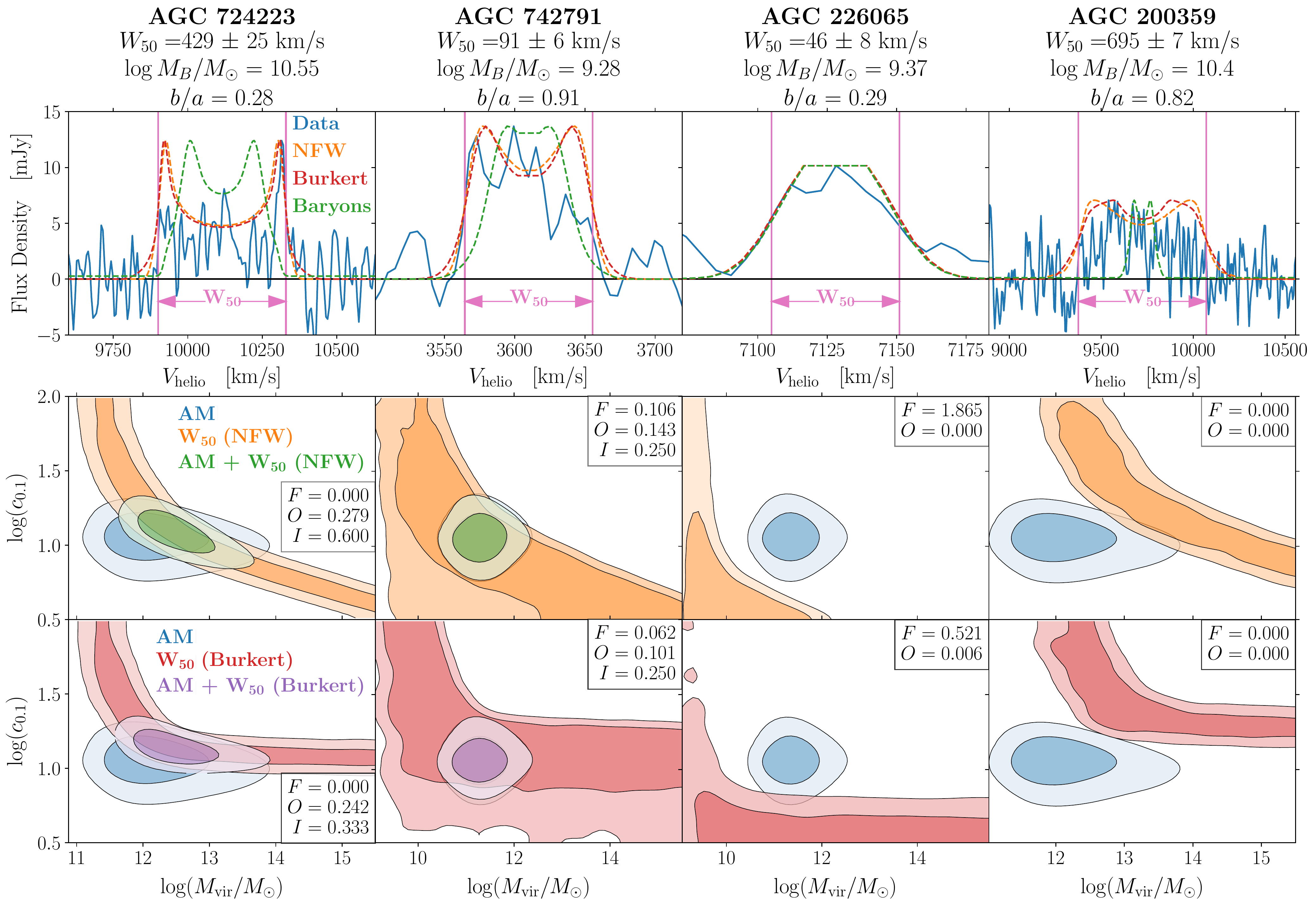} \vspace{-5mm}
\caption{Results for four representative galaxies. From left to right: i) one with a well-constrained DM distribution from kinematics consistent with SHAM, ii) a poorly-constrained distribution consistent with SHAM, iii) an apparently DM-free galaxy and iv) a galaxy with a more massive halo predicted from \lw{} than expected from SHAM. \emph{Top:} The raw ALFALFA spectra, with our best-fit model using an NFW  and Burkert halo profile. We also show the line profile generated by the baryons only, which is calculated using the mean of the priors on the galaxy parameters. As we are only fitting to $W_{50}$, the best fit model is not necessarily a good match to the spectrum, often due to asymmetries. We leave fitting the full spectrum to future work.
\emph{Bottom:} Posteriors on halo mass and concentration from SHAM, kinematics assuming NFW (upper) or Burkert (lower) profiles, and their combination (1 and 2$\sigma$ isoprobability contours shown). The inset lists the $\mathcal{F}$, $\mathcal{O}$ and $\mathcal{I}$ summary statistics (see Table \ref{tab:statistics}).
We only combine the kinematics and SHAM posteriors in the first two cases where they are not in tension.
}\label{fig:grid2d}
}
\end{figure*}

\begin{figure}
\hypertarget{fig:burk_vs_nfw}{%
\centering
\includegraphics[width=8.5cm, height=6.5cm]{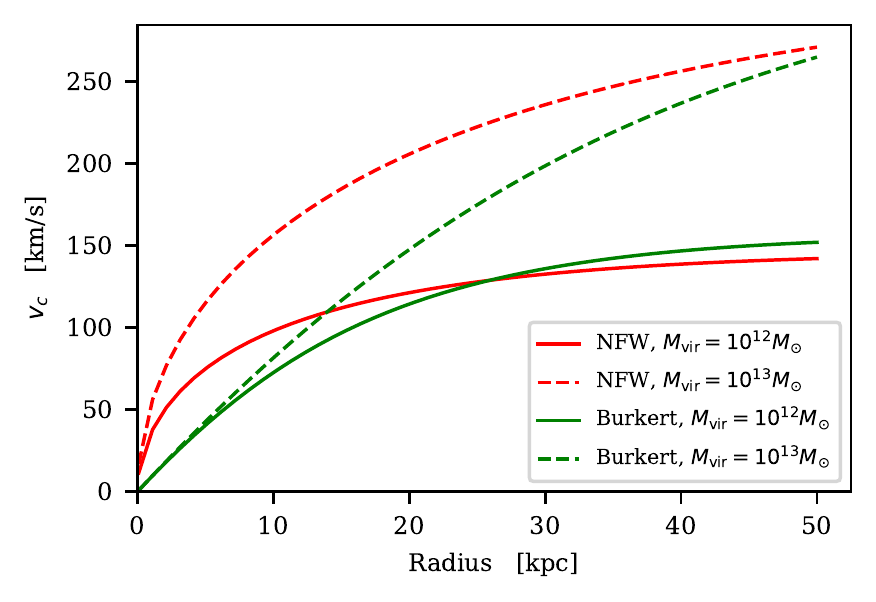} \vspace{-5mm}
\caption{The circular velocity of a dark matter halo with \(c_{0.1}=10\) for difference masses and halo profiles. For the Burkert profile, increasing the halo mass by a factor of 10 results in very little change in circular speed within 20 kpc. For NFW the difference is much more pronounced. This explains why increasing the mass of Burkert haloes that are below a certain concentration only slowly increases the predicted line width, and hence why Burkert masses are poorly constrained.}\label{fig:burk_vs_nfw}
}
\end{figure}

\begin{figure*}
\hypertarget{fig:mbar_mvir}{%
\centering
\includegraphics[width=\textwidth]{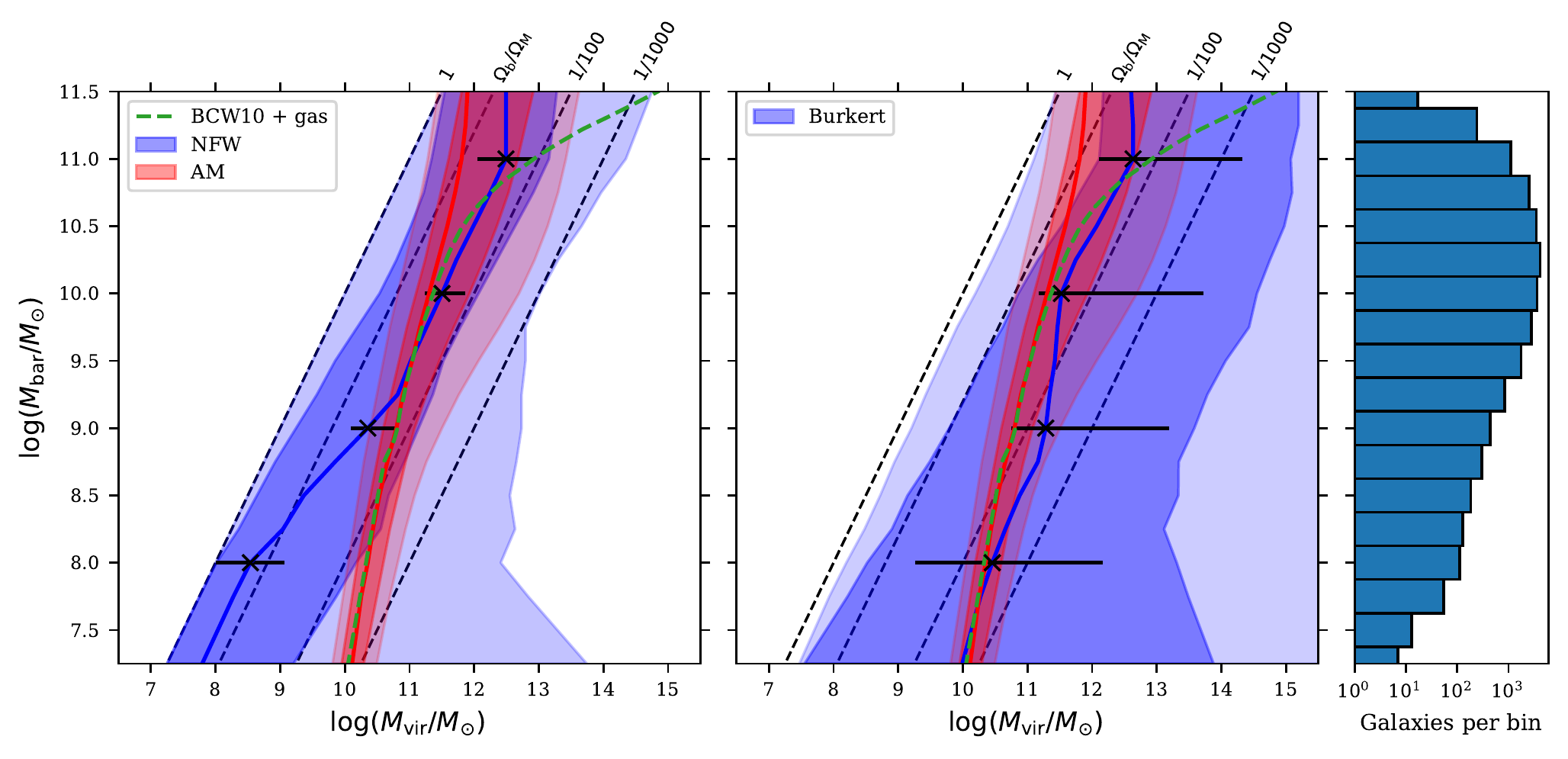}\vspace{-5mm}
\caption{
The baryonic mass--halo mass relation formed by stacking the 1D marginalised posteriors on $\mvir$ in 0.25 dex bins of baryonic mass, for kinematics (blue) and SHAM (red). For kinematics, the mass--concentration prior is applied. The solid line shows the mode of the stacked distribution in each bin, and the shaded regions the 1 and 2$\sigma$ isoprobability contours. `BCW10+gas' shows the best-fit SHAM relationship from \citet{behrooziComprehensiveAnalysisUncertainties2010a}, which is derived for optically-selected galaxies, converted to baryonic mass using the mean gas fraction in each bin for the ALFALFA galaxies.
Black points show the average per-galaxy uncertainties $\mvir$ at given $\mbar$; that these errorbars are much smaller than the width of the blue band shows that this is driven mainly by variations between galaxies in a bin.
The diagonal dashed lines show constant $M{_\text{bar}}/M_{\text{vir}}$ as indicated at the top.}
\label{fig:mbar_mvir}}
\end{figure*}

When we apply the real kinematic model to the whole ALFALFA sample, the resulting 2D posteriors on \(\mvir-c_{0.1}\) vary greatly between galaxies. In Fig.~\ref{fig:grid2d} we show the posteriors for a selection of galaxies to illustrate this diversity, as well as their $\mathcal{F},{O}$ and $\mathcal{I}$ metrics (see Table \ref{tab:statistics} for definitions). The kinematics posterior of AGC 724223 (first panel) is most similar to the theoretical example above. Its \lw{} is large enough relative to $\mbar$ that the baryonic contribution to the dynamical mass within the gas disc is subdominant to DM across the whole range of galaxy priors. Its \(b/a\) is small, so there is a narrow prior on inclination that does not greatly broaden the posterior. It has $\mathcal{F}=0$, because $M_{\text{halo}}=0$ is strongly disfavoured. AGC 742791 (middle right panel) is the opposite case -- the observed \lw{} can be explained with negligible DM within the priors on the baryonic component; it can only have a DM halo of significant mass if \(c_{0.1}\) is so low that most of the halo mass sits outside of the gas disc. This results in much higher values of $\mathcal{F}$.  AGC 742791 is an example between these two extremes. Some region of the priors on galaxy parameters are compatible with having little DM within the gas disc, as shown by the \(2 \sigma\) contour extending down to the baryonic mass of the galaxy. The weak constraints are in large part driven by its high $b/a$, which results in a broad prior on inclination. AGC 200359 (far right panel) also has a high $b/a$, but its \lw{} is very large relative to $\mbar$, so the kinematic constraints are still relatively tight. The posterior strongly disfavours $M_{\text{halo}}=0$, so $\mathcal{F}=0$.

Fig.~\ref{fig:grid2d} also shows the difference between the NFW and Burkert profiles. A higher halo mass is required for the cored Burkert profile to generate the same \lw{} as the cuspy NFW profile. The bend in the banana shape is much more pronounced for the Burkert profile, going almost completely horizontal below a certain concentration for each galaxy, meaning a large range of halo masses generates the same line width at a fixed concentration. The cored Burkert profile has a much more slowly rising circular velocity profile than NFW, so for typical concentrations the larger halo mass does not become apparent until radii larger than the extent of the gas disc. This effect is illustrated in Fig.~\ref{fig:burk_vs_nfw}.

\begin{figure*}
\hypertarget{fig:rfs}{%
\centering
\includegraphics[width=\textwidth]{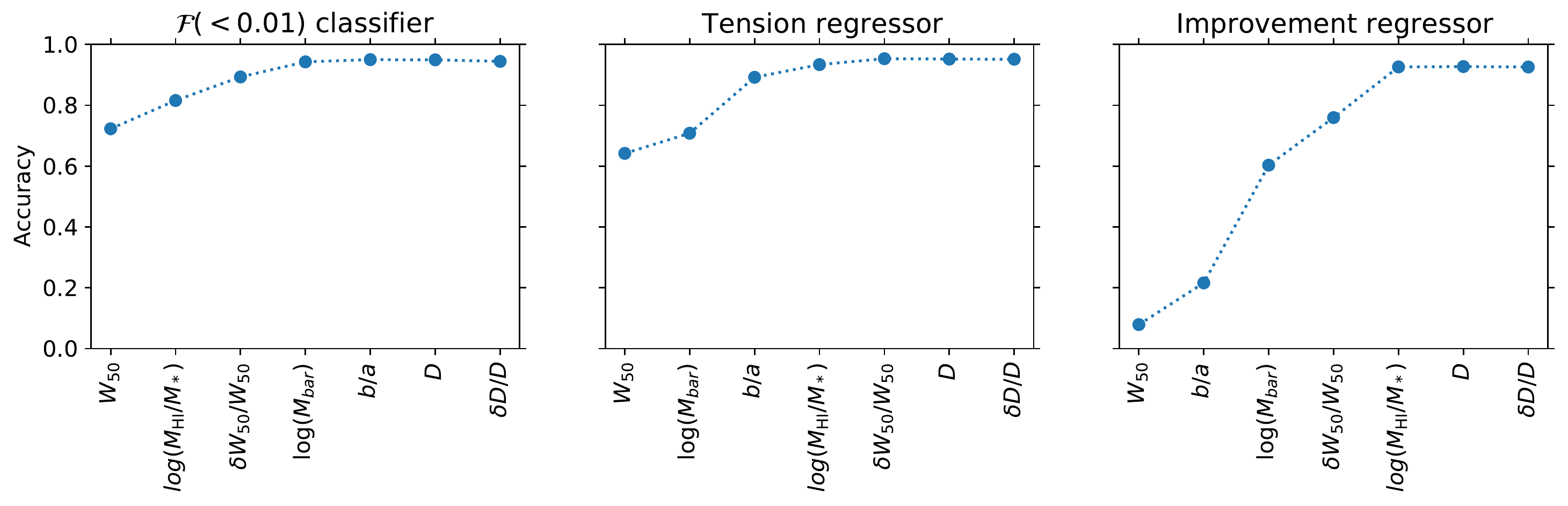}
\vspace{-6mm}
\caption{The test set accuracy of the random forest as a function of the cumulative number of features used to predict the posterior metrics $\mathcal{F}$ (left column), tension $\mathcal{O}$ (middle) and improvement $\mathcal{I}$ (right column) (see Table~\ref{tab:statistics} for definitions). We use random forest regressors for $\mathcal{O}$ and $\mathcal{I}$, and a binary classifier for $\mathcal{F}$ on the condition $\mathcal{F}(< 0.01)$, because its distribution is sharply peaked at $\mathcal{F}=0$. Features are added from left to right in the order that maximises the increment in accuracy, as described in \ref{compatibility-with-mvir0}. Accuracy is measured by the fraction of correctly predicted labels for the classifier, and by $R^2$  (equation~\ref{eq:r2}) for the regressors. We see that \lw{} is the most predictive galaxy property for $\mathcal{F}$ and $\mathcal{O}$, giving reasonable accuracy even when it alone is used to the train the random forest. No galaxy property is predictive on its own for $\mathcal{I}$, but together $W_{50},b/a,\mbar,\delta W_{50}/W_{50}$ and $M_{\text{HI}}/M_*$ yield good accuracy. $D$ and its fractional error add no new information in any case, as would be expected. The results shown are for the NFW profile, but very similar results are obtained with the Burkert profile.}
\label{fig:rfs}
}
\end{figure*}

For all galaxies, the SHAM posterior occupies a constrained region with a defined peak. Towards low mass there is a sharp drop to zero probability, with longer tails to high \(\mvir\), high \(c_{0.1}\) and low \(c_{0.1}\). The SHAM posteriors vary smoothly with $\mbar$, as it is the only galaxy property that the model depends upon (the additional scatter on halo parameters due to the observational uncertainty on $\mbar$ is subdominant to the SHAM model's intrinsic scatter). In particular the posterior smoothly decreases in size in the \(\mvir\) dimension as $\mbar$ decreases (compare the third and fourth columns of Fig.~\ref{fig:grid2d}). This is due to the differing slopes of the BMF and HMF, meaning that the scatter \(\sigma(M_{\text{vir}}|M_{\text{bar}})\) increases towards higher mass, even though the SHAM scatter parameter \(\sigma_{\text{AM}} = (M_{\text{bar}}|\mvir)\) is constant. This also results in an asymmetric posterior: a galaxy of a given $\mbar$ is assigned to a long tail of high mass haloes. We discuss the effect of SHAM scatter further in Section~\ref{sec:am_caveats}. The distribution of \(c_{0.1}\) in the SHAM posterior corresponds to the mass--concentration relationship of haloes that survive the formation time cut, which have a lower mean concentration at a given mass than the whole catalogue.

The galaxies in Fig.~\ref{fig:grid2d} are ordered from least in tension to most in tension. For the first two panels SHAM and kinematics are in good agreement, so \(\mathcal{O}\) is high. Therefore we combine the SHAM and kinematic posteriors, which leads to a significant tightening in the constraints for AGC 724223 due to its tighter kinematic constraint (and wider SHAM constraint due to the higher $\mbar$), whereas there is no improvement for AGC 742791. For AGC 226065 a very low dynamical mass is inferred from \lw, causing tension with AM. The tension is less severe with the Burkert profile, as it is possible to have a larger $\mvir$ whilst keeping the dynamical mass inside the gas disc the same. For cases where SHAM and kinematics are in tension (\overlap<0.01), we do not combine the posteriors.

\hypertarget{population-behaviour}{%
\subsection{Population behaviour}\label{population-behaviour}}

\hypertarget{mbar-mvir-relationship}{%
\subsubsection{\texorpdfstring{\(\mbar-\mvir\) relationship}{\textbackslash mbar-\textbackslash mvir relationship}}\label{mbar-mvir-relationship}}

We now derive a \(\mbar-\mvir\) relationship from our modelling of the \HI{} line width and AM. We do this by stacking the 1D posteriors on \(\mvir\) in bins of \(\mbar\) with 0.25 dex width, applying the mass--concentration prior to break the degeneracy of those parameters. The result is shown in Fig.~\ref{fig:mbar_mvir}. The previously discussed gradual variation in the size and asymmetry of the AM posterior in $\mvir$ as $\mbar$ changes is clearly visible. The kinematic relations have high scatter due to the significant fraction of galaxies for which there is very little constraint on $\mvir$ (e.g. the second panel of Fig.~\ref{fig:grid2d}).
For the Burkert profile the stacked posteriors extend all the way to the upper limit of the prior, due to the cored profile causing a long tail to high mass in most galaxies (e.g. the right panel of Fig.~\ref{fig:grid2d}), which overlaps with the mass--concentration prior.

Despite the large scatter, the mode of the posterior contains important information about the \(\mbar-\mvir\) relationship implied by \HI{} kinematics. It is stable to random resampling for bins with more than \textasciitilde150 galaxies. Above \(\mbar=10^{9.5} \msun\), the NFW and Burkert relationships modes are similar, while towards lower mass the NFW mode diverges from the SHAM mode to a much lower \(\mvir\). The Burkert bends away from the SHAM mode in the other direction, but becomes extremely close to it again for the lowest \(\mbar\) bins. For NFW the slope of the \(\mbar-\mvir\) relationship is much shallower than for AM, and is even compatible with a constant \(\mbar/\mvir\). For NFW, the mean and median of the distribution in each bin are similar to the mode. However due to the long tail to high $\mvir$ for Burkert, the mean and median are different to the mode, and are dependant on the upper bound of the prior. This makes the mode the more robust statistic.

The stacked intervals are very wide for Burkert kinematics, as many individual galaxies have $\mvir$ that are poorly constrained by \lw. The divergence between NFW kinematics and SHAM at low baryonic mass is caused by the substantial population of galaxies with low line widths given their baryonic mass, which seem to have little if any dark matter within the radius probed by their gas disc (e.g. 3$^\text{rd}$ column of Fig.~\ref{fig:grid2d}).

The large scatter of the kinematic relations derive from a combination of per-galaxy uncertainty on $\mvir$ and an offset of the $\mvir$ posteriors of different galaxies in a given $\mbar$ bin. To investigate this, we show in black in Fig.~\ref{fig:mbar_mvir} the per-galaxy uncertainties for four $\mbar$ bins:
the points are at the stacked modal $\mvir$ and the errorbars show the average over all galaxies in the bin of the distance between the mode and the lower/upper $1\sigma$ interval of each individual galaxy posterior. Their size is a much weaker function of $\mbar$ than the stacked intervals, showing that the flaring in the blue band at low $\mbar$ is driven by increased scatter in $\mvir$ \emph{between} galaxies at fixed $\mbar$, rather than increased uncertainties for particular galaxies.

To see how much the relationship is tightened when the galaxies with the weakest constraints are removed, and to check its robustness, we consider a quality cut to the data (Code 1 galaxies only; $\text{SNR} > 10$; $ \delta W_{50}/W_{50} < 0.1$; $b/a < 0.7$, which corresponds to $i\approx 45$ for $q=0.2$), and recalculate \(\mbar-\mvir\) using the remaining 4232 galaxies. The mode and shape of the confidence intervals are unchanged, except for bins with few galaxies after the cut ($M_{\text{bar}} < 10^{8.25} \msun$). The mean width of the $1\sigma$ region is ~30\% smaller for NFW after the cut, but the posterior still reaches the prior bound for Burkert.

\begin{figure*}
\hypertarget{fig:tension_correlation}{%
\centering
\includegraphics[width=\textwidth]{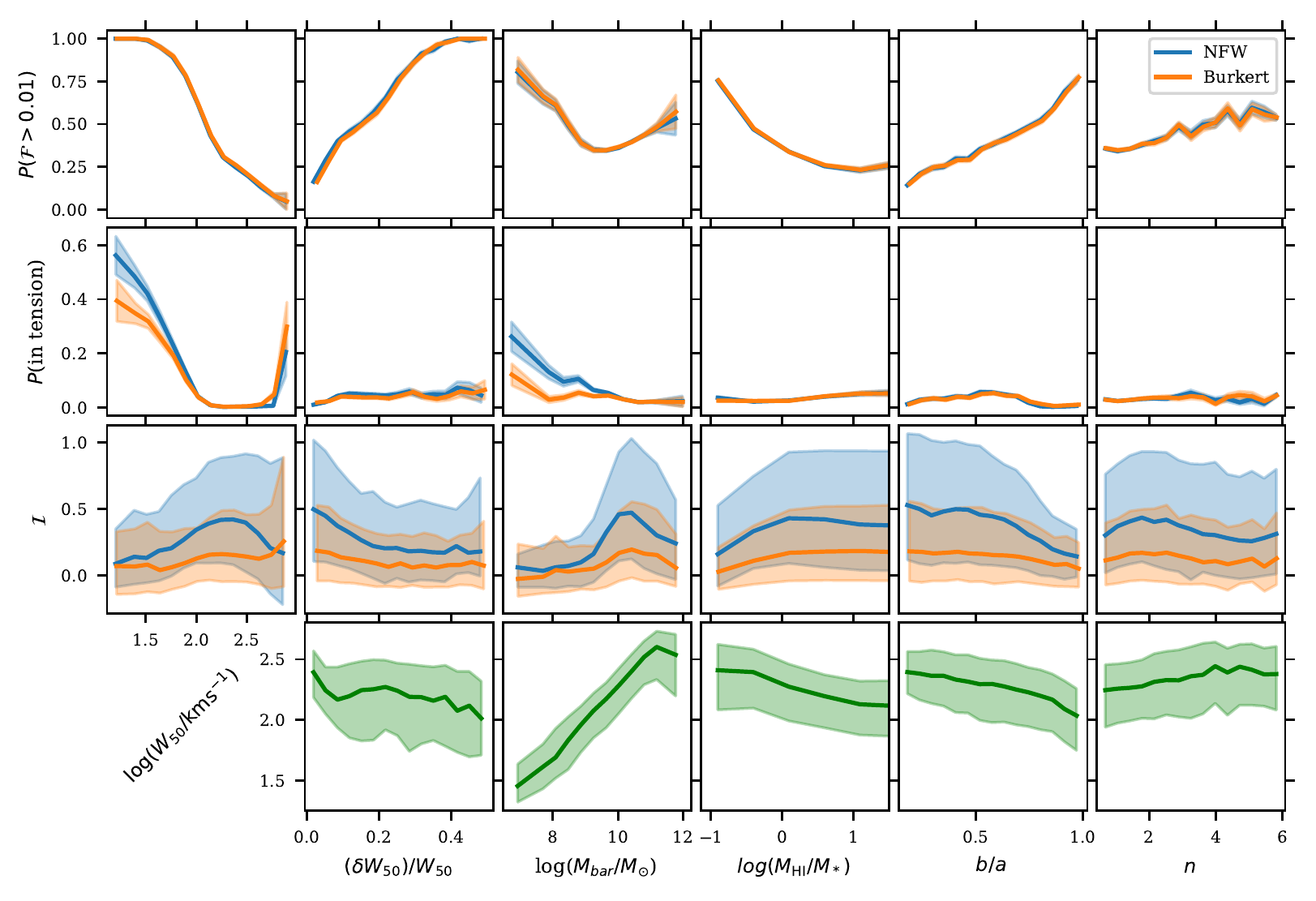}\vspace{-5mm}
\caption{The correlation of galaxy parameters with the $\mathcal{F}$, \overlap{} and $\mathcal{I}$ metrics, and with the ALFALFA line width $W_{50}$. The top two panels show the fraction of galaxies in each bin for which $M_{\text{halo}}=0$ is not excluded by kinematics ($\mathcal{F}>0.01$, top panel) or exhibit tension between SHAM and kinematics ($\mathcal{O} < 0.01$, second panel). The shaded bands show the $1\sigma$ uncertainties calculated by bootstrap resampling each bin. The bottom two panels show the median (solid line) of $\mathcal{I}$ and $\log\w$, with the shaded region showing the $1\sigma$ variation between the galaxies in each bin. We initially set 15 bins in each plot, then merge the outermost bins if there are fewer than 10 galaxies in one of them. We show the correlation of each quantity with $W_{50}$ because we saw in Fig.~\ref{fig:rfs} that this was the most important quantity for determining $\mathcal{F}$, \overlap{} and $\mathcal{I}$. Galaxy parameter values that give tighter constraints (low $\delta W_{50}$, high gas fraction, low $b/a$, high $W_{50}$) produce low values of $\mathcal{F}$ and high values of $\mathcal{I}$. Tension is most prevalent at low $W_{50}$ and low $M_{\text{bar}}$, and is higher for NFW than Burkert.}
}\label{fig:tension_correlation}
\end{figure*}

\hypertarget{compatibility-with-mvir0}{%
\subsubsection{\texorpdfstring{Compatibility with \(\mvir=0\)}{Compatibility with \textbackslash mvir=0}}\label{compatibility-with-mvir0}}

60\% of galaxies in the sample have \(\mathcal{F} < 0.01\) for both Burkert and NFW; the other 40\% have posteriors that are compatible with \(M_{\text{halo}}=0\). To investigate which galaxy properties this depends upon,
we trained a Random Forest Classifier \citep{pedregosaScikitlearnMachineLearning2011} on the sample for the binary classification \(\{\mathcal{F} < 0.01,\mathcal{F} > 0.01\}\), optimising its hyperparameters using 3-fold cross-validation. As our labels are roughly balanced, we assess accuracy using the fraction of correctly predicted labels. To select important features we use the method of \citet[section~3.6]{stiskalekScatterGalaxyhaloConnection2022}, in which galaxy properties are added sequentially to the set of features used to train the random forest. At each increment, the next feature added is the one that generates the greatest improvement in accuracy when combined with the current set of features. This produces a list of features, ordered from most to least important, and the new accuracy after their inclusion. This method of identifying important features avoids ambiguities associated with highly correlated features because the improvement is conditioned on some set of features already being used. The result is shown in the left panel of Fig.~\ref{fig:rfs}. 
For the $\mathcal{F}$ statistic the most important galaxy property is \(W_{50}\), with a much smaller dependence on $\delta  W_{50}/W_{50}$, $\log (M_{\text{HI}}/M_*)$ and $\mbar$.

The top panel of Fig.~\ref{fig:tension_correlation} correlates \(\mathcal{F}\) with various galaxy properties, and the bottom panel correlates these properties with \(W_{50}\). \(\mathcal{F}\) is a strong function of \(W_{50}\), with \(\mathcal{F}>0.01\) for nearly all galaxies in the lowest $W_{50}$ bin. We use the reduced residual method to remove the dependence on $W_{50}$, and find the residuals of $\delta W_{50}/W_{50},\mbar,b/a,n$ correlate positively with $P(\mathcal{F} < 0.01)$, and the residuals of $\log(M_{\text{HI}}/M_*)$ anti-correlate. These are all as expected: higher values of $\delta W_{50}/W_{50}$ and $b/a$ give weaker constraints and therefore higher $\mathcal{F}$; higher gas fractions result in tighter constraints and hence lower $\mathcal{F}$, as a larger $M_{\text{HI}}$ means a larger $r_{\text{HI}}$, so \lw{} probes further into the halo at fixed $\mbar$; higher $\mbar$ galaxies have a lower mean gas fraction and higher $n$, leading to weaker constraints.

\hypertarget{agreement-between-am-and-kinematics}{%
\subsubsection{Agreement between SHAM and kinematics}\label{agreement-between-am-and-kinematics}}

\begin{figure}
\hypertarget{fig:tension_histogram}{%
\centering
\includegraphics{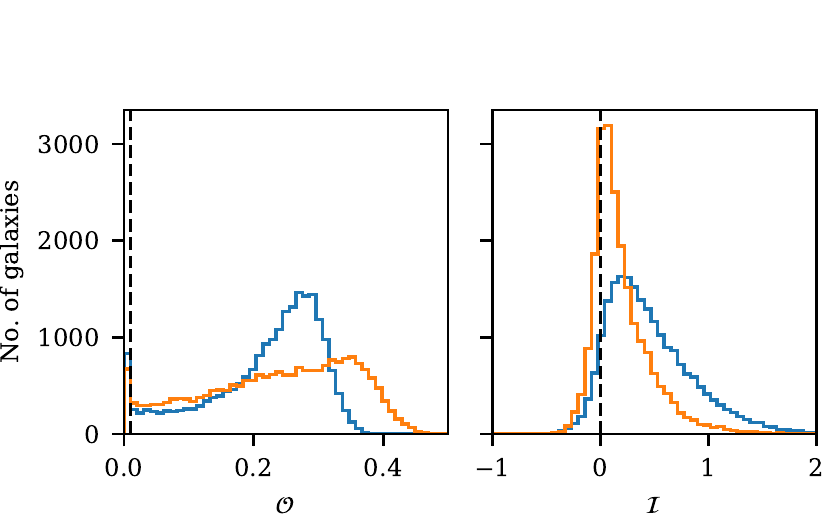}\vspace{-5mm}
\caption{Distribution of the overlap metric $\mathcal{O}$ and improvement metric $\mathcal{I}$ over all galaxies in the sample (see Table \ref{tab:statistics}). SHAM and kinematics are in tension for galaxies with $\mathcal{O}$ to the left of the vertical dashed line. For those with $\mathcal{I}$ to the right of the vertical dashed line, the constraint on $\mvir$ from SHAM is tightened when combined with the constraint from kinematics. For $I=1$ the constraint on $\mvir$ is twice as tight.}\label{fig:tension_histogram}
}
\end{figure}

In Fig.~\ref{fig:tension_histogram} we show the distribution of our tension metric \(\mathcal{O}\) (described in Table \ref{tab:statistics}). Around 1000 galaxies are in tension (\(\mathcal{O} < 0.01\)) for both halo profiles. The second peak in $\mathcal{O}$ corresponds to galaxies where the SHAM posterior lands exactly on the kinematics posterior, such as the left panel of Fig.~\ref{fig:grid2d}. 
Most cases of tension are due to kinematics predicting too little dynamical mass relative to AM, as in the third column of Fig.~\ref{fig:grid2d}. Nearly all these galaxies are also compatible with \(M_{\text{halo}}=0\) according to our \(\mathcal{F}\) statistic. There is a smaller population of galaxies where the dynamical mass inferred from the kinematic model is higher than SHAM (as in the right column of Fig.~\ref{fig:grid2d}).

We correlate tension with galaxy properties in Fig.~\ref{fig:tension_correlation}. The too low/too high dynamical mass populations are visible in the two peaks in the correlation with \(W_{50}\), with a higher peak at low $W_{50}$ than high $W_{50}$. There is higher tension for NFW than Burkert at low $W_{50}$, and also at low $\mbar$, which is also seen in the stacked $\mvir-\mbar$ relationship. There are slightly more galaxies for Burkert that overpredict the dynamical mass than for NFW. It is harder for kinematics to be in tension with SHAM from predicting too high a \(\mvir\) due to the SHAM posterior's long tail to high \(\mvir\), especially at high $\mbar$. 

We apply a random forest regressor to our $\mathcal{O}$ metric to assess its dependence on galaxy properties, once again optimising hyperparameters using 3-fold cross-validation. We use the same importance ranking procedure as before, but this time assessing accuracy using the coefficient of determination

\begin{equation}
R^{2}=1-\frac{\sum_{i}(y_{i,\text{test}}-y_{i,\text{pred}})^{2}}{\sum_{i}(y_{i,\text{test}}-\hat{y}_{\text{test}})^{2}},
\label{eq:r2}
\end{equation}
where $y_{i,\text{test}}$ is the test set value, $y_{i,\text{pred}}$ the corresponding prediction and $\hat{y}_{\text{test}}$ the mean test set value. $R=1$ is perfect prediction, and $R=0$ is given by a model that always predict $\hat{y}_{\text{test}}$, disregarding the input data. We find $O$ is most dependant on \(W_{50}\), and to a lesser extent on $\mbar$ and $b/a$ (Fig.~\ref{fig:rfs}, middle panel).

\begin{figure*}
\hypertarget{fig:mbar_w50_hist2d}{%
\centering
\includegraphics[width=\textwidth]{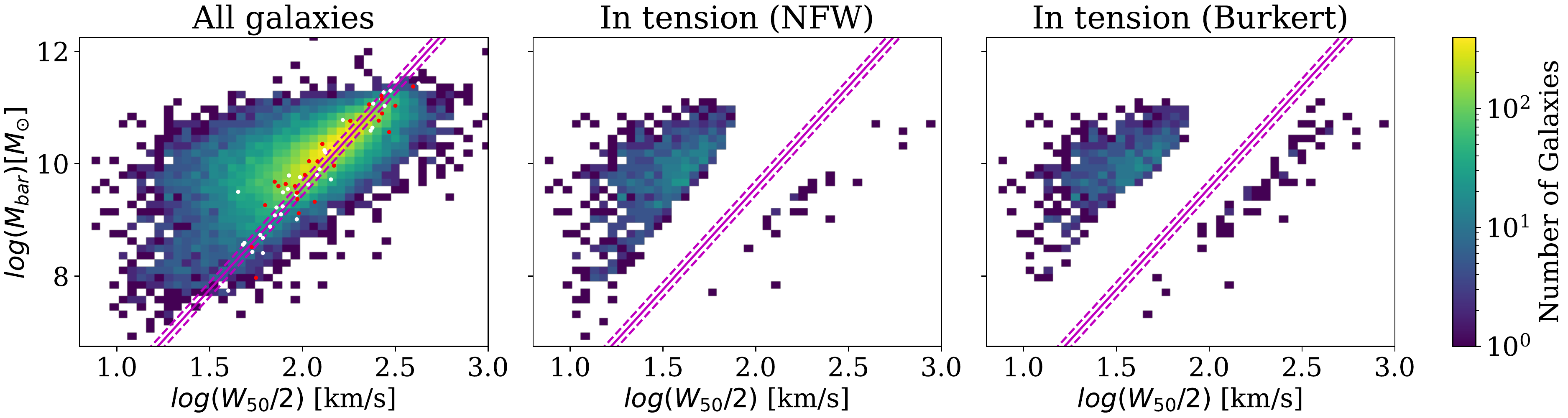}\vspace{-5mm}
\caption{A heat map of the distribution of \(W_{50}\) and \(\mbar\) for the whole ALFALFA sample,  and for galaxies where SHAM and kinematics are in tension. The $W_{50}$ have been corrected for inclination assuming the intrinsic relative thickness $q=0.2$. The \(\mbar-W_{50}\) relationship from SPARC \citep{lelliBaryonicTullyFisherRelation2019} is shown (magenta line) with its $3\sigma$ intrinsic scatter (dashed lines). The 25 galaxies that are in SPARC$\times$ALFALFA$\times$NSA are plotted both using their ALFALFA data (red dots) and their SPARC data (white dots) in the left panel. For the latter we use the $W_{\text{m50}}$ line width measure and kinematic inclinations, as these were used to derive the SPARC \(\mbar-W_{50}\) relationship.  } \label{fig:mbar_w50_hist2d}
}
\end{figure*}

In Fig.~\ref{fig:mbar_w50_hist2d} we plot the 2D \(W_{50}-\mbar\) distribution of the ALFALFA sample, which shows clearly that tension primarily occurs when \(W_{50}\) is low for a given \(\mbar\). The split is very clean in the \(\log W_{50}-\log \mbar\) plane and the \(\mbar-W_{50}\) relationship from the SPARC sample lies within it. The scatter in the ALFALFA data is much greater than in SPARC. We see from the relatively similar scatter of the ALFALFA and SPARC data describing the same galaxies (red and white dots) that the ALFALFA data is reasonably accurate for the SPARC sample. Therefore the increased scatter is due to systematics in the ALFALFA$\times$NSA data that are not present for galaxies that are also in SPARC and/or because the ALFALFA$\times$NSA sample has a different intrinsic \(\log W_{50}-\log \mbar\) distribution to SPARC e.g. we do not expect the entire sample to be rotationally supported and in equilibrium. We discuss potential systematics extensively in Section~\ref{sec:kinematic_caveats}.

We repeat the random forest regressor procedure for the improvement statistic $\mathcal{I}$, and find that it is dependant on a greater number of features than $\mathcal{F}$ and $\mathcal{O}$ (Fig.~\ref{fig:rfs}, right panel). The improvement $\mathcal{I}$  is much greater for NFW than Burkert due to the Burkert posterior's long tail to high mass, which for many galaxies aligns with the high mass of tail of the SHAM posterior. We correlate $\mathcal{I}$ with galaxy properties in Fig.~\ref{fig:tension_correlation}. The positive correlation of $\mathcal{I}$  with $\mbar$ is because the SHAM constraint is much weaker at high $\mbar$ due to the increasing scatter \(\sigma(\mvir|\mbar)\). The correlations with other galaxy parameters are for the same reasons as discussed for the $\mathcal{F}$ statistic, as $I$ is strongly correlated with the tightness of the kinematic constraint.

\vspace{-5mm}

\hypertarget{discussion}{%
\section{Discussion}\label{discussion}}

\hypertarget{interpretation-of-the-results}{%
\subsection{Interpretation of the results}\label{interpretation-of-the-results}}

\subsubsection{Abundance matching vs. kinematic constraints}

Our \lw-based kinematic model and our abundance matching (AM) model produce Bayesian posteriors on the mass and concentration of individual galaxies. The constraints offered on halo mass by kinematics are reasonably strong (<1 dex with the mass--concentration prior applied) for around 40\% of galaxies when assuming a cuspy profile. When assuming a Burkert profile the constraints are much weaker, often only providing a lower limit on mass, as the more slowly rising circular velocity profiles due to the core mean that haloes with vastly different masses can have relatively minor differences in rotational velocity within the gas disc. 
The model is flexible enough that the mode of the posterior produces a $W_{50}$ within $2\sigma$ of the observed value for all galaxies. Thus it is not possible for the constraints to be artificially tightened by a poor fit, which is a concern when fitting resolved rotation curves.

In general the constraints offered by SHAM are tighter than for kinematics. When assuming an NFW profile, combining SHAM with kinematics yields stronger constraints on halo mass for a majority of galaxies, with a 1-sigma constraint that is twice as tight for around 1/5 of the sample. The improvement is more pronounced at the high mass end, where the SHAM constraints are weaker. Our analysis therefore shows that SHAM can be augmented with the abundant \HI{} line widths of future large scale surveys to constrain better the distribution of dark matter, especially for haloes with cuspy profiles.

The improvement from combining probes may in fact be even stronger, as the tightness of our SHAM constraints are likely overestimated because our analysis does not account for the uncertainties on the SHAM model itself (e.g. in the proxy and scatter). This is especially true at the low-mass end, where there is no clustering data to constrain the SHAM parameters. We discuss the uncertainty on the SHAM model further in Section~\ref{sec:am_caveats}.

\subsubsection{Tension}

Overall we do not find significant tension between kinematics and SHAM when comparing the posteriors of individual galaxies, apart from a population of galaxies for which the line width predicts too small a dynamical mass relative to AM, and a smaller population of galaxies for which the inverse is true. The lack of tension is to some extent unsurprising given the weak constraints on halo properties from kinematics for many galaxies. For our SHAM model the mean $\mvir$ changes by only 2 dex over 4 dex in baryonic mass, which is comparable to the scatter in $\mvir$ from kinematics for a substantial fraction of our sample even with the mass--concentration prior applied.

Although more individual galaxies are in tension between SHAM and kinematics for both profiles at lower baryonic mass (in part due to the tighter constraints from SHAM at low mass), fewer are in tension with Burkert than NFW. This is also clearly visible in the stacked \(\mbar-\mvir\) relationship: the mode of the stacked posterior lies at much lower \(\mvir\) when assuming an NFW profile, compared to both SHAM and the Burkert profile, which are in good agreement with each other. Therefore both individual and stacked posteriors suggest that at low mass NFW underpredicts halo mass relative to AM/Burkert. This is expected when fitting a cuspy profile to a cored halo \citep{trujillo-gomezEmergenceDarkMatterdeficient2022}. Therefore the disagreement may imply that core formation is a decreasing function of mass, and/or that SHAM needs further refining for HI-selected samples.

There is much literature, both theoretical and observational, on dark matter core formation (see \citealt{delpopoloReviewSolutionsCuspCore2021} for a review). A popular hypothesis is that core formation proceeds through the kinematic heating of the dark matter in galactic centres by supernova driven cycles of gas inflows and outflows \citep{readMassLossDwarf2005,pontzenColdDarkMatter2014}. Some simulations \citep[e.g.][]{dicintioDependenceDarkMatter2014} and analytic calculations \citep{penarrubiaCOUPLINGCORECUSP2012} have found core formation to be a function of \(M_*/\mvir\), with \(M_*\) a proxy for the energy available to drive gas outflows with supernovae, and \(\mvir\) related to the amount of DM that needs to be removed and the depth of the potential well that it needs to be removed from. \citet{readDarkMatterHeats2019} find observational evidence for this in local dwarf galaxies. From our stacked $M_*-\mvir$ relationship (which is qualitatively very similar to our $\mbar-\mvir$ relationship), we see no such correlation with \(M_*/\mvir\): SHAM is most in agreement with cored halos relative to cusps at low $M_*$, rather than at some value of \(M_*/\mvir\), as suggested by the above models. However our results in this regard are weak, and do not rule out any scenario. The disagreement between kinematics with NFW and SHAM at low mass could also be resolved by haloes expanding to lower concentrations than in DMO simulations. Semi-analytical studies based on SHAM have found some evidence that this is necessary to explain the Tully--Fisher \citep{desmondTullyFisherMasssizeRelations2015} and radial acceleration \citep{desmondStatisticalInvestigationMass2017} relations. Both these papers use a relaxation model in which the halo as a whole can expand or contract relative to its pristine NFW profile \citep{duttonRevisedModelFormation2007}, finding mild evidence for expansion. However, applying a different relaxation model to the radial acceleration relation, \citet{paranjapeRadialAccelerationRelation2021} found evidence for expansion only in the outer halo.

The \(\mbar - W_{50}\) relationship for the SPARC galaxy sample \citep{lelliBaryonicTullyFisherRelation2019} is extremely tight (see Fig.~\ref{fig:mbar_w50_hist2d}). \citet{papastergisAccurateMeasurementBaryonic2016} take an extremely restricted sample of only 90 ALFALFA galaxies and construct a similarly tight relationship: their criteria are edge-on, high gas fraction and \(W_{50} > 100 \kms\), and they inspect each spectrum for residual noise. They find a correlation between increasing offset of a galaxy from their mean relationship to lower \(W_{50}\) and the degree to which the \HI{} spectrum only has a single peak (measured by kurtosis). The offset also increases at lower baryonic mass. Our full ALFALFA$\times$NSA sample is a very different population of galaxies compared to these two studies, so it is expected that the \(\mbar - W_{50}\) distribution will differ. However with such a large sample based on unresolved observations, some of our galaxies will inevitably have biased halo properties due to systematic uncertainties. Galaxies that lie far from the literature relations fall under particular suspicion. We consider potential systematics in the kinematic modelling in detail in Section~\ref{sec:kinematic_caveats}, but we note here that our results are robust to basic quality cuts on SNR and $b/a$, such as the one used to test the $\mbar-\mvir$ relationship in Section~\ref{mbar-mvir-relationship}.

\subsubsection{\(\mbar-\mvir\) relationship}

A key result is the \(\mbar-\mvir\) relationship derived from \HI{} kinematics (Fig.~\ref{fig:mbar_mvir}), which we find to be roughly linear in log-space. This is reminiscent of the baryonic Tully--Fisher relation (BTFR) between $M_\text{bar}$ and $V_\text{rot}$, which is linear over six decades in mass. Canonical optically-selected SHAM relationships such as that of BCW10, however, display a break in the stellar mass--halo mass relation (SHMR) around the Milky Way mass. Generating a linear BTFR from such SHAM relationships is non-trivial \citep{desmondScatterResidualCorrelations2017}. Observations of massive spirals have also failed to detect the SHMR break expected from SHAM \citep{liComprehensiveCatalogDark2020,postiDynamicalEvidenceMorphologydependent2021,mancerapinaImpactGasDisc2022}, and \citet{mcgaughDarkMatterHalo2021a} pointed out that the halo mass predictions for the Milky Way and Andromeda from kinematics lie well below the SHAM prediction. Neither the \(M_*-\mvir\) or \(\mbar-\mvir\) relationship from our kinematic model display a break, but continue approximately linearly with a mode close to the cosmic baryon fraction at the highest masses. This result is robust to quality cuts in the data. Although for the Burkert profile the relationship is very wide, the modal relation is still linear.

\citet{postiDynamicalEvidenceMorphologydependent2021}, also applying the mass--concentration relationship as a prior in their kinematic analysis, argue for a linear SHMR relationship for late-type galaxies, with elliptical galaxies displaying the expected break. Unlike \citeauthor{postiDynamicalEvidenceMorphologydependent2021}, we do not have a clean sample of late-type galaxies. When assessed by standard cuts on colour and S\'{e}rsic index (\(n\)) it appears the majority of galaxies in our sample above \(10^{10.5} \msun\) are early-type. When we split our sample using a crude cut on morphological type (\(n\) \textless{} 2 for late-type, \(n>3\) for early-type) we do not find differing relationships in \(\mbar-\mvir\). Multiple empirical studies using different methods have attempted to measure the SHMR for passive and star forming galaxies separately, but no consensus has yet been reached on whether they differ \citep{wechslerConnectionGalaxiesTheir2018}.

Our fiducial SHAM model, based on $\HI{}$-selected galaxies, has a scatter of $\sigma_{\text{AM}}=0.42\,\text{dex}$, whereas the scatter for optically-selected galaxies is well constrained to \textasciitilde0.2 dex (at least under the assumption that $\sigma_{\text{AM}}$ is not a function of mass; e.g.~\citealt{reddickConnectionGalaxiesDark2013}). As $\sigma_{\text{AM}}$ increases the highest mass galaxies are increasingly assigned to the much more numerous lower mass haloes, washing out the characteristic break in the mean \(\mbar-\mvir\) relationship, as can be seen in Fig.~\ref{fig:mbar_mvir}. However there is still an increasing scatter \(\sigma(M_{\text{vir}}|M_{\text{bar}})\) towards high mass, with the posterior extending to high $\mvir$. There is also still a break in galaxy formation efficiency as a function of halo mass. The fiducial model is in better agreement with our kinematic relationship than BCW10, but we caution that $\sigma_{\text{AM}}$ is very poorly constrained for \HI{}-selected galaxies. The model also assigns around a third of the most massive galaxies into haloes such that $\mbar/\mvir> \Omega_{\text{b}}/\Omega_{\text{M}}$, suggesting SHAM models with scatter this high may not produce realistic galaxy populations, even if they reproduce the clustering signal.

\hypertarget{sec:kinematic_caveats}{%
\subsection{Kinematic modelling caveats}\label{sec:kinematic_caveats}}

\hypertarget{halo-mismatch-concentration}{%
\subsubsection{Baryonic effects on halo density profiles}\label{halo-mismatch-concentration}}

We have studied the NFW and Burkert profiles as representative cases of a cusped and cored profile respectively. However previous studies have shown that no halo profile is a good fit to all observations \citep{katzTestingFeedbackmodifiedDark2017,liComprehensiveCatalogDark2020}. This may be explained by the difficulty in calculating baryon-induced modifications to haloes, due to the uncertainty in the numerical implementation of baryonic physics and resolution constraints. Initially haloes are expected to contract adiabatically from baryonic infall as galaxies form (\citealt{liIncorporatingBaryondrivenContraction2022b} incorporate this effect into their halo fitting procedure). The subsequent expulsion of gas from galaxies by star-formation is then expected to expand the halo. Recently \cite{velmaniPreprint2022a} studied halo relaxation in large volume hydrodynamical simulations and found it to vary substantially with halo mass and concentration, and star formation rate. \citet{paranjapeDistributionHIVelocity2021a} demonstrated that changes to halo relaxation physics can significant alter \lw.

Any baryonic effect that modifies halo density profiles would be expected to alter our mass and concentration constraints, and thus our analysis would need to be repeated for halo profiles inspired by hydrodynamical simulations (e.g.~\citealt{dicintioDependenceDarkMatter2014}). As there is currently no convergence in the precise effects of baryons on dark matter across a range of galaxy scales, this is left for future work. Ultimately, theoretical and observational progress in understanding the net effect of feedback processes is required before the modification of haloes by baryons can be robustly accounted for in kinematic analyses.

\hypertarget{rotation-curve-not-reflected-in-line-width}{%
\subsubsection{Modelling the line width}\label{rotation-curve-not-reflected-in-line-width}}

Our model assumes the line width is the product of a galaxy's azimuthally-averaged rotation curve and \HI{} surface density, with a \HI{} velocity dispersion of 10km/s. We tested this using the SPARC sample and found good agreement. However very few of these galaxies had observed line widths \(< 100 \kms\), which is the region in which most of the tension is found. Furthermore we see a trend in Fig.~\ref{fig:model_residuals} where towards lower line width there is weaker agreement between model and observations, although this could be explained by uncertainty in the extrapolation of RCs. 

Pressure support has the effect of reducing the rotational velocity below the circular velocity, causing the dynamical mass to be underestimated \citep{bureauEnvironmentRamPressure2002,ohHIGHRESOLUTIONMASSMODELS2015,iorioLITTLETHINGS3D2016a}. This "asymmetric drift" correction becomes important in galaxies where $V_{\text{rot}}$ is comparable to the gas dispersion, and becomes more important at larger radii. Our models for $\sigma_{\text{HI}}$ and $\Sigma_{\text{HI}}$
do not capture the galaxy-to-galaxy variation required to sensibly apply the correction, which is challenging even with resolved data. As a simple test of the sensitivity of our results to asymmetric drift, we rerun our inference replacing the assumption that $V_{\text{rot}}(r) = V_{\text{c}}(r)$ with $V_{\text{rot}}^2(r) = V_{\text{c}}^2(r) - \sigma_{\text{HI}}^2$, with $\sigma_{\text{HI}}=10$km/s. We find that for our stacked $\mbar-\mvir$ relationship the mode of $\mvir$ is increased by 0.5 dex for NFW and 0.15 dex for Burkert in the lowest mass bin ($\mbar=10^{7.25} \msun$), where the effect is greatest. At $\mbar=10^{8.5} \msun$ the difference with our fiducial model is less than 0.15 dex for both haloes. We conclude that, although there is an increasing number of galaxies towards lower mass and \lw{} that are affected by asymmetric drift, our overall conclusions are likely robust to it. 

Non-equilibrium motions are also expected to be increasingly important towards lower mass, due to supernovas driving gas out of the plane of the disc, and radial outflows (e.g. \citealt{verbekeNewAstrophysicalSolution2017}). Asymmetries due to the increasing irregularity of galaxies may also cause \lw{} not to reflect the dynamical mass \citep{reynoldsAsymmetriesLVHISVIVA2020}. High velocity clouds in the observed galaxy can also create high velocity wings in the flux profile, leading to \lw{} overpredicting the rotational velocity \citep{schulmanSurveyHighVelocityClouds1994}.

\vspace{-4mm} 

\hypertarget{line-width-measurement}{%
\subsubsection{Measuring the line width}\label{line-width-measurement}}

Extracting the line width from often noisy spectra is a difficult process. We investigate the difference between the base ALFALFA catalogue \(W_{50}\) line width measurement and the \(W_{\text{Yu85}}\) measurement from the \citet{yuStatisticalAnalysisProfile2022} reanalysis. \(W_{\textrm{Yu85}}\) implies a slightly lower mass for most galaxies, but there is a significant population for which it infers a much higher mass. The strength and trends of tension with galaxy properties are the same for both line width measures, although there is not good agreement on which specific galaxies are in tension. \(W_{\textrm{Yu85}}\) produces a very similar stacked \(\mbar-\mvir\) relationship to $W_{50}$, with the mode not different by more than 0.4 dex in any bin.

\citet{haynesAreciboLegacyFast2018} caution that at low SNR, radio frequency interference can cause the line width to be underestimated. This could potentially explain the population of galaxies with extremely low \lw{} relative to their baryonic mass. However we find that galaxies with \HI{} code 2 are not overrepresented among in-tension galaxies, and we do not find a trend between SNR and tension above $\text{SNR}=6$, suggesting this cannot be the sole cause. \citet{yuStatisticalAnalysisProfile2022} provide a different cut on galaxies more vulnerable to RFI. Again we found these galaxies were not overrepresented among the in-tension galaxies. Another potential source of error is confusion, where the separation between two galaxies is smaller than the beam width. \citet{jonesWhenStackingConfusing2016} show that the impact of confusion for the catalogue as a whole is not significant, although they say it is easy to identify specific examples. Using the flag for crowding in the reanalysis of \citet{yuStatisticalAnalysisProfile2022}, we found that crowded galaxies are not over-represented among the in-tension galaxies. 

\cite{ballGeneralistAutomatedALFALFA2022} present an ALFALFA reanalysis that utilises a similar curve-of-growth based algorithm and crowding analysis to \citeauthor{yuStatisticalAnalysisProfile2022}, with the aim of minimising the BTFR scatter by removing outliers. Future work may benefit from testing their methods and sample selections in the mass-modelling context.

\hypertarget{inclination}{%
\subsubsection{Inclination}\label{sec:caveat_inclination}}

As discussed in Section~\ref{sec:inclination}, the gas disk inclination calculated from the observed optical axis ratio can be highly inaccurate. The most troubling potential inclination systematic is when the measured $b/a$ is biased low, leading to too high (edge-on) an inclination, too low an inferred intrinsic \lw{} and hence too low a dynamical mass within the gas disc. The majority of our in-tension galaxies are of the type where the dynamical mass inferred from \lw{} is lower than the SHAM expectations, with the frequency of tension increasing at lower $\mbar$. In our SPARC sample we find a single severe underestimation of $b/a$ in the optical, for a heavily barred face-on galaxy. As bars are most prevalent in galaxies with stellar masses \(10^9 < M_* / \msun < 10^{11}\) \citep{mendez-abreuWhichGalaxiesHost2010}, this is unlikely to explain the increasing tension that we observed towards even lower $\mbar$. 

Several irregular galaxies in the SPARC sample had a NSA $b/a$ that was somewhat too low (given both their kinematic inclination and visual appearance), which we accounted for by adopting a $10\%$ error on $b/a$. This sample was too small to look for a correlation between mass or flux and disagreement with kinematic inclination, to see if the effect increases towards lower mass, as does the tension. In general we expect irregularity to increase towards low $\mbar$, as the weaker self-gravity of the system makes it more susceptible to internal and environmental effects.

Applying a flat prior on $q$, when the true distribution for the sample is likely peaked at $q\approx0.2$, will cause the inferred masses to be biased low. We calculated the \(\mbar-\mvir\) relationship for the $q=0.2$ inclination model, and found it made little difference in the lowest and highest mass bins, where it raises $\mvir$ somewhat. However it is in these bins that we expect $q$ to be higher than 0.2.

Finally, \citet{almeidaTriaxialityCanExplain2019} have argued that low dynamical masses may be caused by ignoring triaxiality when calculating inclination using the optical $b/a$. We also neglect triaxiality, so our results are subject to the same potential bias. A more sophisticated analysis could use the expected population distribution of inclination to infer the distribution of axial ratios as a function of galaxy properties \citep{putkoInferring3DShapes2019}. However this is complicated for ALFALFA, as the selection function of blind spectroscopic \HI{} surveys is dependant on inclination \citep{langFirstResultsJodrell2003}.

\hypertarget{baryonic-mass}{%
\subsubsection{Baryonic mass}\label{baryonic-mass}}

Using an erroneously high $\mbar$ will cause $\mvir$ to be underestimated. The uncertainties on $M_{\text{HI}}$ are smaller than those on $M_*$. We verified the NSA $M_*$ in two ways. Firstly we compared it to the GALEX survey and found consistency within the uncertainties. Secondly we compared the NSA to SPARC, and also found good agreement on the whole, but with increasing disagreement towards low mass (\(< 10^9 \msun\)). However low mass galaxies tend to be more gas-dominated, so the bias on $\mvir$ from $M_*$ will be less important, suggesting it is unlikely to be the cause of the observed trend of tension with mass. \cite{ballGeneralistAutomatedALFALFA2022} also identified foreground stars as a potential source of overestimated stellar masses, finding this pathology in 11\% of the most extreme BTFR outliers in their data.

\HI{} self-absorption has not been accounted for in \HI{} masses. Depending on the model used, the correction can range from insignificant except for the most edge-on galaxies, to a 30\% correction for all galaxies (see \citealt{jonesALFALFAMassFunction2018} for a thorough discussion). Underestimating the \HI{} mass causes the halo circular velocity to be overestimated, and the size of the \HI{} disc to be underestimated. Both of these result in too high a halo mass being inferred. Applying the inclination-based correction from \citeauthor{jonesALFALFAMassFunction2018} ($\Delta \log M_{\text{HI}}=0.13\log (b/a)$) results in insignificant differences to our results above $\mbar=10^9\msun$. However at lower mass, where galaxies are more gas-rich, the effect is significant, with the mode of the stacked $\mvir$ distribution 0.4 dex lower at $\mbar=10^{7.5}\msun$ for NFW, putting more galaxies in tension. As the \citeauthor{jonesALFALFAMassFunction2018} model is based on thin disc galaxies, the true effect may be larger for thicker dwarf galaxies. Future work may benefit from improved modelling of \HI{} self-absorption.

\hypertarget{gas-distribution}{%
\subsubsection{Gas distribution}\label{gas-distribution}}

The adopted gas model from \cite{wangNewLessonsHI2016} is based on a sample of 500 dwarf and spiral galaxies, with masses down to \(M_{\text{HI}} = 10^7 \msun\). They find that early-type galaxies, although still lying on the \(M_{\text{HI}}-D_{\text{HI}}\) relation, tend to have flatter gas profiles, with a larger fraction of their gas lying outside of \(D_{\text{HI}}\). Adopting too low an $r_{\text{HI}}$ causes $\mvir$ to be overestimated. On the other hand, explaining the tension between SHAM and kinematics at the low mass end would require that the true size of the gas disc be smaller than in the model. The \(M_{\text{HI}}-r_{\text{HI}}\) relationship of W16 shows no evolution towards lower mass. The most in-tension galaxies in this regime, for which the baryonic mass alone is enough to generate the observed line width, are little sensitive to changes in gas distribution as both the stars and gas are contained within the gas disc.

\hypertarget{abundance-matching-caveats-secam_caveats}{%
\subsection{Abundance matching caveats }\label{sec:am_caveats}}

The abundance matching relationship for gas-selected galaxies is constrained by clustering only for \(\log M_{\text{B}} > 9.4\), and even for those ranges it is poorly constrained (see fig.~14 of ST21) due to the weaker clustering of \HI-selected samples and the comparatively small sample size of \HI{} surveys.

Rather than sample from the entire posterior of ST21, which includes regions with extremely high \(\sigma_{\text{AM}}\) for which the galaxy--halo connection is essentially fully randomised, we adopted their maximum likelihood point under the assumption that the SHAM parameters are independent of mass (although this was strongly ruled out for optically-selected samples by ST21). As discussed previously, the high mass end is particularly sensitive to \(\sigma_{\text{AM}}\). We find far weaker dependence on the $z_{\textrm{cut}}$ parameter. 

The extrapolation of the baryonic mass function becomes important for galaxies with \(\log \mbar < 10^{8.5} M_{\odot}\). However the trends in our results are not noticeably different in the extrapolated regime. The only way to increase the baryon fraction for SHAM at the low mass end (hence reducing tension with kinematics) is to strongly increase the steepness of the baryonic mass function.

Finally, at the faint end it is possible that a qualitatively different SHAM prescription is required to deal with the different formation scenarios of lower mass galaxies. ST21 showed that fainter optically-selected samples require higher scatter, and argue that low mass galaxies may require a different set of SHAM assumptions, such as increased galaxy formation bias or a difference between satellite and central galaxies. On the other hand, \citet{nadlerMilkyWaySatellite2020} find an upper limit $\sigma_{\text{AM}}=0.2$ for Milky Way satellites.

\hypertarget{comparison-to-literature}{%
\subsection{Comparison to literature}\label{comparison-to-literature}}

Previous studies, most notably of the SPARC galaxies, have used resolved rotation curves to infer halo properties \citep{katzTestingFeedbackmodifiedDark2017,liComprehensiveCatalogDark2020}. The advantage of resolved rotation curves is that they are able to provide much more information on density profiles than a summary statistic that effectively samples the RC at a single radius.
\citet{liComprehensiveCatalogDark2020} tested a wide variety of rotation curves, and found that in general cored profiles such as Burkert provided better fits than cuspy profiles such as NFW, with many galaxies favouring cores even at higher mass. 
Even with the full rotation curve, the constraints on halo properties are weak in some cases. For example \citet{liComprehensiveCatalogDark2020} find many galaxies for which the 1$\sigma$ halo mass constraint spans over two orders of magnitude. However for many of the galaxies they recover a good constraint without any prior applied, as the shape of the rotation curves is enough to break the degeneracy between mass and concentration. Our approach is complementary, sacrificing precision of the dynamical measurements for a much larger sample size and thus trading potential systematic error in relating the small SPARC sample to the entire halo population for weaker galaxy-by-galaxy constraints. For us, robust conclusions are available only statistically across the full sample. 

Our analysis sheds light on the ``small-scale problems'' of $\Lambda$CDM \citep{bullockSmallScaleChallengesCDM2017}. The Too-Big-to-Fail problem  (TBTF) is the observation that the kinematically-inferred halo masses of the Milky Way's satellites are much lower than the masses of the largest subhaloes of Milky Way-sized haloes in N-body simulations \citep{boylan-kolchinTooBigFail2011b}. 
This was later generalised to populations of field galaxies that were found to have kinematics that implied a lower halo mass than predicted by abundance matching \citep{ferreroDarkMatterHaloes2012}. \citet{papastergisThereTooBig2015} studied the problem in a sample of ALFALFA isolated dwarfs for which resolved rotation curves also exist, using a SHAM procedure that ranks galaxies by their line width. Fitting an NFW profile, they found that haloes with resolved outer rotational velocities of less than 25km/s are incompatible with the haloes implied by AM. They find that fitting a halo profile with a mass-dependant core reduces the tension, but does not fully alleviate it. Although we do not separate satellite and field galaxies, or restrict our sample to galaxies with resolved rotation curves (and hence more robust kinematic halo masses), our results are similar to the above studies: we also find a population of dwarf galaxies for which the halo mass inferred from kinematics is significantly below the SHAM prediction, such that the two measurements are in statistical tension. The disagreement is partially alleviated by fitting a Burkert profile instead of NFW. There is much literature on proposed solutions to TBTF for both satellite and field galaxies, including modelling and observational uncertainties, and new dark matter physics (see \citealt{papastergisAssessmentTooBig2016} for a review). 

A related small-scale problem is the observed diversity of rotation curves at fixed galaxy mass, which does not appear in $\Lambda\text{CDM}$ simulations \citep{omanUnexpectedDiversityDwarf2015}. Baryonic models which solve small scale problems such as TBTF through core-formation create cores too uniformly, and therefore fail to generate this rotation curve diversity \citep{salesBaryonicSolutionsChallenges2022a}. In our sample we observe great diversity in $W_{50}$ at fixed $\mbar$ at lower mass, where the gas disc does not probe so far into the halo. This could be indicative of different arrangements of the baryons and/or different DM central densities at fixed halo mass, assuming the abundance matching relation does not flare at low mass. However, as we do not have accurate baryonic distributions we cannot provide more concrete results.

We find a significant number of ALFALFA galaxies are dark matter-deficient according to our model, as their line width is completely explained by the baryons alone. The existence of apparently dark matter-deficient galaxies has been previously noted for the ALFALFA sample by \cite{guoFurtherEvidencePopulation2020b}, using a simple method where the dynamical mass is estimated from the gas disc scale length and the observed line width without full modelling. They apply quality cuts to the sample and find 19 dark-matter deficient galaxies (14 of which are isolated) out of a sample of 324 (although \citealt{almeidaTriaxialityCanExplain2019} argue this may due to neglected triaxiality, see Section~\ref{sec:caveat_inclination}).

\cite{mancerapinaBaryonicTullyFisherRelation2019} study a sample of six ALFALFA galaxies with low linewidths for their baryon masses using HI interferometric data, deriving resolved RCs (with 2-3 resolution elements per galaxy side) which support the galaxies being baryon-dominated. Furthermore, using higher resolution observations, \citet{mancerapinaNoNeedDark2021} identified an apparently ``dark matter-free'' isolated galaxy in the sample, although the inclination is still a potentially significant systematic uncertainty and its stability in the absence of dark matter has been contested \citep{sellwoodUltradiffuseGalaxyAGC2022}. Two dark matter-free dwarf galaxies have also been identified using globular cluster dynamics (\citealt{vandokkumGalaxyLackingDark2018,vandokkumSecondGalaxyMissing2019}, although see \citealt{saifollahiNumberGlobularClusters2021}) speculated to have formed from gas stripped in a galaxy--galaxy collision \citep{vandokkumTrailDarkMatterfree2022}. 

In general, claimed observational detections of dark-matter deficient galaxies tend to be controversial due to modelling uncertainties, even with far better data than our unresolved observations. It is interesting to speculate, however, whether the galaxies we identify as being plausibly dark matter-deficient would remain so given more precise measurements. \citet{jacksonDarkMatterdeficientDwarf2021} and \citet{morenoGalaxiesLackingDark2022a} find dark matter-free galaxies produced in tidal interactions in simulations, the latter predicting that 30\% of central galaxies host at least one dark matter-free satellite.

Another approach to comparing \HI{} kinematics with \(\Lambda\text{CDM}\) expectations is to forward model the line width velocity function using either semi-analytical models \citep{chauhanVelocityFunctionTest2019a,paranjapeDistributionHIVelocity2021a} or hydrodynamical simulations \citep{duttonNIHAOXVIIDiversity2018a,el-badryGasKinematicsFIRE2018}. This does not require the inclinations of individual galaxies, avoiding a major source of uncertainty. \citet{duttonNIHAOXVIIDiversity2018a} find that the velocity function for dwarf galaxies in the hydrodynamical NIHAO simulations are in good agreement with ALFALFA line widths in the range \(10 < W_{50}/2 < 80 \kms\). They identify turbulent motions, projection effects due to intrinsic \HI{} disc thickness and flattened DM distributions as important factors in lowering the observed line widths relative to expectations. We do not properly account for turbulent motions (although we test our model sensitivity to them) and we assume an infinitely thin \HI{} disc. We find less tension at low line width when fitting a Burkert profile, but our inferred core-formation dependence is in disagreement with the $M_*/\mvir$ dependence seen in their simulation. It is plausible that the turbulent motions, \HI{} disc thickness and feedback physics in their simulation account for the differences with our results.

\citet{duttaDarkMatterHaloes2022} use an abundance matching-based method to extract $M_{\text{HI}}$--$\mvir$--$V_{\text{rot}}$--\lw{} scaling relations for the ALFALFA sample. They use group finder-based halo masses to obtain a reduced halo mass function corresponding to the ALFALFA sample, which they abundance match to the ALFALFA \HI{}MF. The resulting \HI{}-selected \HI{}-to-halo mass relationship (their fig. 7) is similar to the mean of our stacked abundance matching \HI{}-to-halo mass relationship at low mass, but at high mass has a much stronger break. This is largely driven by the lack of AM scatter in their model (compared to 0.42 dex in ours), which prevents an apples-to-apples comparison.

\vspace{-4mm}

\hypertarget{future-work}{%
\subsection{Future work}\label{future-work}}

This paper presents a first attempt to compare the halo properties inferred from abundance matching and \HI{} line widths for an entire population of HI-selected galaxies.  Future \HI{} surveys will improve the constraints on SHAM models for \HI{}-selected galaxies by reducing the uncertainty on galaxy clustering and extending it to lower masses, allowing a more robust assessment of the agreement between the two methods and their relative constraining power.

The increased precision and reduction in systematics on \lw{} of future surveys should also improve the constraints from kinematics. The increasing number of observed line widths will also increase the statistical power at the low and high mass ends. Future surveys will also allow both \HI{} line width and abundance matching studies to be extended to higher redshifts. \citet{ponomarevaMIGHTEEHBaryonicTully2021} have already used line widths from MeerKAT to study the BTFR out to z=0.081. \citet{glowackiRedshiftEvolutionBaryonic2021} predict evolution in the BTFR with redshift from the SIMBA simulation.

More information is contained in the \HI{} flux profile, of which \lw{} is a summary statistic. Exploiting this would increase the precision of halo parameter inference.
Using a similar model as this work, \citet{paranjapeDistributionHIVelocity2021a} showed this
by performing a full spectrum fitting for some nearby galaxies with very well resolved ALFALFA spectra. This method is potentially very powerful if it can be applied to whole populations of galaxies. A potential problem is the uncertainty in the detailed \HI{} distribution, which may have cores, holes or asymmetries. These may bias the inferred halo properties if not adequately modelled.

Finally, other methods of inferring the properties of DM haloes could potentially be combined, which could probe the DM distribution at different radial distances from the halo centre. The \HI{} line width probes the central region of the halo, but weak lensing measures the acceleration towards the outskirts of stacked galaxies. The velocity dispersions of stars in early-type galaxies (or of galaxies in groups or clusters) could also be used. For example, \citet{schulzTestingAdiabaticContraction2010} used weak lensing to measure the dark matter halo profile in the outskirts of massive elliptical galaxies, extrapolated it to the centre assuming an NFW halo and then compared the resulting central dynamic mass to the SDSS velocity dispersion, finding evidence for halo contraction.

\vspace{-3.5mm}

\hypertarget{conclusions}{%
\section{Conclusions}\label{conclusions}}

We have compared the constraints on halo mass and concentration inferred from the kinematic modelling of the \HI{} line width with those inferred from an (inverse) abundance matching model specifically tailored to HI-selected galaxies, for the \textasciitilde22,000 galaxies in the ALFALFA$\times$NSA data set. Our conclusions are as follows:
\begin{itemize}
\tightlist
\item
  The two methods produce consistent halo constraints galaxy-by-galaxy in most cases, with the kinematics posterior broader and requiring a mass--concentration prior for bounded constraints on either quantity.
 \item
  The halo posteriors of SHAM can be augmented with information from the \HI{} line width to produce tighter constraints on the dark matter distributions of individual galaxies. The gains are greater when assuming a cuspier halo profile.
\item
  Towards low baryonic mass there is an increasing population of galaxies with smaller line widths than expected from abundance matching. For some galaxies this implies a dynamically insignificant amount of dark matter within their gas disc, leading to extremely high baryon fractions when the halo is extrapolated to the virial radius. The disagreement with abundance matching is more severe when fitting an NFW halo than Burkert, which we interpret as weak evidence for a cored central DM density at low baryonic mass. There is a smaller population of galaxies for which SHAM and kinematics disagree because the dynamical mass inferred from kinematics is higher than from AM. 
\item
 The $M_\text{bar}-M_\text{vir}$ relation reconstructed from \HI{} kinematics is in statistical agreement with that from SHAM (Fig.~\ref{fig:mbar_mvir}). It is however closer to a power-law, with a deviation (especially assuming an NFW profile) towards lower $M_\text{vir}$ at fixed $M_\text{bar}$ at the faint end. When assuming a Burkert profile there is less information to be gleaned on $M_\text{vir}$ from the line width, resulting in a very uncertain relation.
\item 
  We formulate statistics to quantify whether a galaxy i) exhibits tension between its kinematics and SHAM modelling results, ii) affords a strong improvement in halo constraints by combining the two methods, and iii) has $M_{\text{halo}}=0$ excluded by the kinematic modelling. We also develop a machine learning-based method for assessing the extent to which these statistics correlate with various galaxy properties, finding line width to be the most important feature in each case.
\end{itemize}

Our analysis demonstrates the potential for combined photometric and spectroscopic constraints on the galaxy--halo connection, even when using low-resolution spectroscopic products such as \HI{} line widths. With future surveys set to improve dramatically our knowledge of the \HI{} universe, we anticipate that our framework will be useful for inferring DM distributions, constraining kinematic and empirical models, and advancing understanding of the physical processes that underlie galaxy formation.

\section*{Acknowledgements}

We thank Anastasia Ponomareva, Richard Stiskalek, Harley Katz, Martin Rey and Stacy McGaugh for useful inputs and discussions, and Federico Lelli, Richard Stiskalek, Martha Haynes and Jing Wang for sharing their data with us.

HD was supported by a Junior Research Fellowship at St John's College, Oxford, a McWilliams Fellowship at Carnegie Mellon University and a Royal Society University Research Fellowship (grant no. 211046).

This project has received funding from the European Research Council (ERC) under the European Union’s Horizon 2020 research and innovation programme (grant agreement No 693024).

\section*{Data Availability}

Data from the ALFALFA and NSA surveys are publicly available at \url{http://egg.astro.cornell.edu/alfalfa/data/index.php} and \url{https://www.sdss.org/dr13/manga/manga-target-selection/nsa/}. The SPARC database is located at \url{http://astroweb.cwru.edu/SPARC/}. 
The data of the \citet{yuStatisticalAnalysisProfile2022} reanalysis of ALFALFA will be publicly released with their paper.
Data from the Uchuu simulation suite is available at \url{http://skiesanduniverses.org/Simulations/Uchuu/}.
Other data will be made available on reasonable request to the corresponding author.





\bibliographystyle{mnras}
\bibliography{new} 


\bsp	
\label{lastpage}
\end{document}